\documentclass[structabstract]{aa}
\usepackage[varg]{txfonts}
\usepackage{graphicx}
\usepackage{natbib}

\usepackage{subfig}

\usepackage{latexsym}

\usepackage{multirow}

\begin{document}

\title{HAWK-I infrared supernova search in starburst galaxies
\thanks{ESO proposal: 083.D-0259, 085.D-0335, 085.D-0348, 087.D-0494, 087.D-0922. GTC proposal: GTC50-11B}
}
\author{M. Miluzio\inst{1}, E. Cappellaro\inst{2}, M.T. Botticella\inst{3},  G. Cresci\inst{4}, L. Greggio\inst{2},  F.  Mannucci\inst{4}, S. Benetti\inst{2}, Bufano, F.\inst{5}, Elias-Rosa, N.\inst{6},  A. Pastorello\inst{2}, M. Turatto\inst{2}, Zampieri, L. \inst{2}}

\institute{Department of Astronomy, Padova University,
               Vicolo dell'Osservatorio 3, I-35122, Padova, Italy\\
               \email{matteo.miluzio@unipd.it}
    	\and
		INAF,  Osservatorio Astronomico di Padova, vicolo dell'Osservatorio 5, Padova,
		 35122 Italy
	\and 
		INAF, Osservatorio Astronomico di Capodimonte, Salita Moiariello 16, Napoli, 80131 Italy
	\and
		INAF,  Osservatorio Astrofisico di Arcetri,  Largo Enrico Fermi 5, Firenze,
		50125 Italy
	\and
		Departamento de Ciencias Fisicas, Universidad Andr\'es Bello, Av. Rep\'ublica 252, Santiago, 
		Chile
	\and
           	Institut de Ci\'encies de l'Espai (IEEC-CSIC), Facultat de Ci\'encies, Campus UAB, Bellaterra, 
	08193 Spain
	}
	
\date{Received: ????; Revised: ??????; Accepted: ????? }
\titlerunning{HAWK-I  SN search}
\authorrunning{M. Miluzio et al.}

\abstract
{The use of SN rates to probe  explosion scenarios and to trace the cosmic star formation history received a boost from a number of synoptic surveys. There has been a recent claim of a mismatch by a factor of two between star formation  and core collapse SN rates, and different explanations have been proposed for this discrepancy. }
{We  attempted an independent test of the relation between star formation and supernova rates in the extreme environment of starburst galaxies, where both star formation and extinction are extremely high.}
{To this aim we conducted an infrared supernova search in a sample of local starburts galaxies. The rational to search in the infrared is to reduce the bias due to extinction, which is one of the putative reasons for the observed discrepancy between star formation and supernova rates. To evaluate the outcome of the search we developed a MonteCarlo simulation tool that is used to predict the number and properties of the expected supernovae based on the search characteristics and the current understanding  of starburst galaxies and supernovae.  }
{During the search we discovered 6 supernovae (4 with spectroscopic classification) which is in excellent agreement with the prediction of the MonteCarlo simulation tool that is, on average, $5.3\pm2.3$ events.}
{The number of supernovae detected in starburst galaxies is consistent with that predicted from their high star formation rate  when we recognize that a major fraction ($\sim 60\%$) of the events remains hidden in the unaccessible, high density nuclear regions because of a combination of reduced search efficiency and high extinction.}


\maketitle

\section{Introduction}\label{intro}

The rate of supernovae (SNe) is a key quantity in astrophysics that provides a crucial test for  stellar evolution  theory and an input for the modeling of  galaxy evolution with direct impact on the  chemical enrichment and the feedback mechanism.
Core-collapse SNe  (SN~CC), because of their short-lived progenitors, trace the current star  formation rate (SFR).  Conversely, for an adopted  SFR, measurements of the SN~CC rates give information on the mass range of their progenitors as well as  the slope of 
the initial mass function at the high mass end.
SN~Ia, resulting from the thermonuclear explosion of a white dwarf in 
a binary system, show a wide range of delay times from star formation to explosion. Therefore, the SN~Ia rate reflects the long-term star formation history of the parent stellar system. 
Recently, it has been claimed that a significant fraction of SN~Ia have a short delay time, 
possibly as short as $10^7$ years \citep{mannucci:2006zi}. Like for  CC SN, the rate of such {\it prompt} SN~Ia events is  expected to be proportional to the current SFR.  

In one of the early attempts to compare the SN and SF rates, \cite{cappellaro:1999dg} found that the SN~CC rate in galaxies with different $U-V$ color matches the predicted SFR when adopting a mass range $10$ M$_\odot < M < 40$ M$_\odot$ for the SN~CC progenitors. 

In the last decade there was a enormous improvement in the measurement of the cosmic SFR with the careful combination of many different probes \citep[eg.][]{hopkins:2006wj}. A most relevant feature is that the SFR reaches a maximum  at a redshift $z\sim 1$ and hereafter begins
 to decrease down to the current rate which is over one order of magnitude lower than at peak. 

A significant effort was also devoted to the measurement of the cosmic SN rate: although much of the focus was for type SN~Ia, a few estimates of the SN~CC rates were also published both for the local Universe \citep{li:2011qf} and at high redshifts \citep{dahlen:2004on,cappellaro:2005qb,botticella:2008fr,bazin:2009fd,graur:2011ys,melinder:2012ly,dahlen:2012zr}. While the new local SN~CC rate confirms previous results, with a much better statistics and lower systematic errors, the evolution with redshift was found to track very well the SFR evolution, considering the large uncertainties in the extinction corrections. Again, to best match the observed SN and SF rates it was argued that the lower limit for SN~CC progenitor had to be $\sim 10$  M$_\odot$ \citep{botticella:2008fr,blanc:2008gr}.

At about the same time, following a different line of research, the analysis of archival images allowed the identification of the precursors for a number of nearby SN~CC. From the often very scanty but precious photometry, and using stellar evolution models, one can  estimate the SN precursor mass. The uncertainties are in general quite large, as confirmed from the discrepancy in the mass estimates from different groups, but this analysis suggests a lower limit for SN~CC progenitors of $8 \pm 1$ M$_\odot$ \citep{smartt:2009mq}. If this value is adopted, the observed SN rates would result a factor two smaller than those expected from the observed SFR. This was identified by some authors as a "SN rate problem" (e.g. \citealt{horiuchi:2011xv}.
While one should remind that the uncertainties on SFR rate calibrations are still large \citep{botticella:2012sh,kennicutt:2012pt}, it also true that there is a number of possible biases in the SN rate estimates. The two most severe are the possible underestimate of a large population of faint SN~CC and/or the underestimate of the correction for extinction \citep{horiuchi:2011xv,mattila:2012er}.

In particular, \cite{mannucci:2007tz, cresci:2007ly} and, more recently, \cite{mattila:2012er} argued that a significant fraction of SN~CC remains hidden in the nuclear region of starburst galaxies, with a loss of up to  $\sim$70-90\% in the highly dust-enshrouded environments of (ultra-)luminous infrared galaxies( U/LIRGs). This effect is expected to be more important at high redshift because of the larger fraction of starburst galaxies. Indeed, when a correction for this hidden SN fraction is included in the rate calculation     the discrepancy between SN and SF rates at high redshifts seems to disappear (\citealt{melinder:2012ly,dahlen:2012zr}; the ''missing fraction'' correction adopted in these works was from \cite{mattila:2012er}. It is currently unclear if this effect is large enough to explain also the discrepancy observed in the local Universe with somewhat conflicting evidences from the statistics of SNe in the Local Group galaxies \citep{botticella:2012sh,mattila:2012er} and large sample SN searches \citep{li:2011qf}.

Entering in this debate, we planned for an infrared SN search in a sample of local starburst galaxies (SBs). The idea was to verify the link between SN and SF rates in an environment where  star formation is very high, 1-2 order of magnitude  higher than in normal star-forming galaxies. By observing in the  K-band we were aiming to reduce the bias due to extinction (A$_{\rm K}\sim 0.1$A$_{\rm V}$). 

The idea is not new.  A first attempt of a dedicated SN search in SBs was performed in the optical band by \cite{richmond:1998bl}. During the  search  only a handful of events were detected leading the authors to conclude that the rate of (unobscured) SNe in SBs is the same as in quiescent galaxies. A similar conclusion was reached by \cite{navasardyan:2001fk}, again based on optical data.
As for infrared SN search, after a few unsuccessful attempts  \citep{grossan:1999uq,bregman:2000kx}, the first results of a systematic search in SBs were reported by \cite{maiolino:2002vn} and \cite{mannucci:2003wl}. They found that  the observed SN rate in SBs was indeed one order of magnitude higher then expected for the galaxy blue luminosities but still 3-10 times lower than would be expected from the far infrared (FIR) luminosity. Among the possible explanation for the remaining discrepancy, they suggested extreme extinction in the galaxy nuclear regions (A$_{\rm V}>25$mag), which would dim SNe even in the near-IR, and  insufficient spatial resolution to probe the very nuclear regions. 
The reliability of the use of NIR search for obscured SNe in the nuclear and circumnuclear regions of  active starburst galaxies was also investigated by \cite{mattila:2001gf} taking into account in particular the problem of extinction. They conclude that with a modest investment of observational time it may be possible to discover a number of nuclear  SNe. 
A negative search  for transients in NICMOS images retrieved from the Hubble Space Telescope archive suggests that the same biases likely affect also space-based, high spatial 
resolution observations \citep{cresci:2007ly}.

The same approach was used by \cite{mattila:2007nb} but with ground based, adaptive optics (AO) assisted observations. The  application of this technique led to the discovery of a handful of SNe  \citep{kankare:2008ys,kankare:2012zr} but not yet to an estimate of the SN~CC rate.

Until now, about a dozen SNe have been discovered by IR SN searches, not all with spectroscopic confirmation. The number is higher if we include also events first detected in the optical and re-discovered by the IR searches. Therefore the statistics is still very low and many of the original questions are still unanswered. This gave us the motivations to make a new attempt exploiting the opportunity offered by HAWK-I, the infrared camera mounted at the ESO VLT telescope. 

The paper is divided in two parts:  the first part describe the observing program, namely
the galaxy sample and the search strategy  in Sect. \ref{galaxysample}, the data reduction 
in Sect.~\ref{dataredu}, the SN discoveries and classification in Sect.~\ref{sndisc} while in Sect.~\ref{artstar} we detail the procedure to estimate the search detection efficiency. 
The second part is devoted to the description of  a simulation tool  which is used to predict, based on our current knowledge of SBs properties and on the specific features of our SN search, the number of {\em expected} SN detections (Sect.~\ref{montecarlo}).
Finally, we compare the number and properties of the expected and observed events (Sect.~\ref{results}) and draw our conclusions (Sect.~\ref{summary}).

Throughout this paper we assume the following cosmological parameters:  $H_0 = 72\, {\rm km}\,{\rm s}^{-1}\,{\rm Mpc}^{-1}$, $\Omega_\Lambda = 0.73$ and $\Omega_M = 0.27$.

\section{The SN search program}

\subsection{Galaxy sample}\label{galaxysample}

Starbursts are galaxies with very high star formation rate, of the order of 10-100 M$_{\sun}\,$ yr$^{-1}$ compared to the few M$_{\sun}\,$ yr$^{-1}$  of normal star forming galaxies in the local universe. Given that in a typical galaxy  the very high SFR will rapidly consume the gas reservoir, it is thought that the starburst is a temporary phase in the galaxy evolution. The fact that many SBs are in close pairs or have disturbed morphologies point to the interaction as a  dominant, although possibly not unique, reason of the phenomena \citep{gallagher:1993fk}. The ultra-violet radiation from young, massive stars heats the surrounding dust and is re-emitted in the far infrared. Indeed the most luminous SBs in the local Universe are LIRGs , with {\bf $11<\log (L_{IR}/{\rm L}_{\sun}) < 12 $}, and ULIRGs, with $\log (L_{IR}/{\rm L}_{\sun}) >12\,$  \citep{sanders:1996fl}. 

For our project we selected from the IRAS Revised Bright Galaxy Sample \citep{sanders:2003uq} a sample of SBs with total infrared (TIR) luminosity $\log (L_{TIR}/{\rm L}_{\sun})>11 $\/  and redshift $z<0.07$.  With the additional requirement that the targets are accessible from Paranal in the  April to September observing season (to fit in one of the ESO allocation period) we retrieved a sample of 30 SBs.

\begin{figure}
\centering
\includegraphics[width=0.5\textwidth]{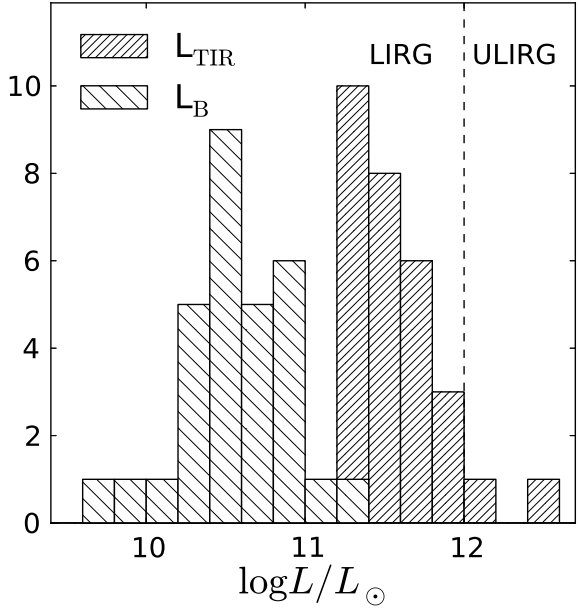}
\caption{Distribution of the B and FIR luminosities for the SB galaxies of our sample.} \label{ltirlb}
\end{figure}

The list of SBs is reported in Tab.~\ref{galaxies}. Along with the galaxy name and equatorial coordinates (cols. 1-3) we report the heliocentric redshift (col. 4),   $\log L_{TIR}$ and  $\log L_B$ (cols. 5 and 6; cf. Sect.~\ref{TIRdef}),   the Hubble type (col. 7), the SFR and the expected SN rates  (cols 8, 9) derived from $L_{TIR}$ as described in Sect.~\ref{TIRdef}.  Galaxy data have been retrieved from NED\footnote{The NASA/IPAC Extragalactic Database (NED) is operated by the Jet Propulsion Laboratory, California Institute of Technology, under contract with the National Aeronautics and Space Administration.}.  In the last column we listed (in boldface) the designation of the SNe  discovered in our search which are the basis for our analysis. For completeness we also list (in italics) the SNe discovered by other SN searches outside our monitoring period.  
The distribution of $L_B$ and $L_{TIR}$  are compared in Fig.~\ref{ltirlb} showing that, as typical for SBs,  $L_{TIR}$ is on average a factor ten higher than $L_B$, whereas for normal star forming galaxies  $L_{TIR} \sim L_B$. We notice that almost all galaxies are LIRGs and only two are ULIRGS. Most  galaxies of the sample are isolated  ($\sim 60-70$\%) while the remaining are  double/interacting galaxies or contain double nuclei,  signature of a recent merger.  Several galaxies of the sample are asymmetrical, disturbed, or show warps, bars and tidal tails.

\subsection{Search strategy}\label{searchstrategy}
To search  SNe in the selected SB sample we used the  HAWK-I  instrument installed at the ESO VLT telescope at Cerro Paranal (Chile).  HAWK-I is a NIR ($0.85-2.5\, \mu$m ) wide-field imager with a mosaic of  four  Hawaii-2RG detectors.  The  total field of view is $7.5\arcmin \times 7.5\arcmin$  with a scale of $0.106\, \arcsec/$pix. Even in poor seeing conditions ($>1.5$ arcsec)
the instrument allows  to achieve ${\rm S/N}\sim10$ for a K=20 magnitude star with  a 15 min exposure.

The infrared light curves of SNe evolve relatively slowly, remaining within one/two magnitudes from maximum for two/three months \citep{mattila:2001gf} and therefore  an IR SN search does not require frequent monitoring. We planned for an average of three visits per galaxy per semester, for a total of 80-100 visits.  The monitoring campaign was scheduled in service mode and we did not set  tight constraints for the sky conditions. This and the relatively short duration of the observing blocks made the program well suited as filler.  We notice that we had no influence on the actual scheduling of the observations which followed the rules of the ESO  service mode scheduler. 

Eventually, the fraction of useful observing time was 100\% of the allocated time  in the first season, and 70\% in the second and third semesters. The log of the observations is reported in Tab.~\ref{log} where for each galaxy we list the epoch of observations (MJD),   the seeing (FWHM in arcsec), and the minimum and maximum magnitude limit for SN detection across the image (cf. Sec.~\ref{artstar}). In total, we obtained 210   K-band exposures (exposure time 15min), with an average of about 3 visits per galaxy per semester. Because of the time loss,  three galaxies were not  monitored in the last two seasons.

It turned out that  the average image quality was quite good:  for $\sim 90$ \% of the exposures the seeing was  less than $1.0\arcsec$, with an average $FWHM$ across the whole program  of $0.6 \arcsec$.
  
\begin{table*}
\begin{centering}
\caption{The SB galaxy sample. The last column report our 6 SNe (in bold) with other SNe discovered previously in the galaxy sample. }\label{galaxies}
\small\addtolength{\tabcolsep}{-2pt}
\begin{tabular}{lcclcccccc}
\hline\hline
galaxy          &  R.A. & Dec. & redshift & $\log L_{TIR}$ & 
$\log L_{\rm B}$ & Hubble &SFR                   &SN rate      &SNe\\
  designation              &        \multicolumn{2}{c}{J2000.0}                 &    &   [L$_{\sun}$]        &     [L$_{\sun}$]          &    type    &[M$_{\sun}\,{\rm yr}^{-1}$]   &[SN yr$^{-1}$]&\\
\hline 
CGCG011-076       &11 21 13.3        &-02 59 08 & 0.025    &11.28   & 10.35 & 2.9&32.1&0.38&       \\ 
CGCG043-099       &13 01 49.9        &+04 20 01  & 0.037    &11.59   &10.51 &3.4&65.4&0.77 &                         \\ 
ESO148-IG002      &23 15 46.6        &-59 03 14 & 0.045    &11.94   &10.82&7.9&148.8&1.75  &                     \\
ESO239-IG002      &22 49 39.6        &-48 51 01 & 0.043    &11.75   &10.88&-0.1&95.7&1.13 &                           \\ 
ESO244-G012       &01 18 08.6        &-44 27 40 & 0.023    &11.32   &10.20&5.3&35.7&0.42  &                       \\ 
ESO264-G036       &10 43 07.0        &-46 12 43 & 0.023    &11.24   &10.70&1.5&29.1&0.34  &                     \\ 
ESO286-IG019      &20 58 27.4        &-42 38 57 & 0.043    &11.95   &11.13&10.0&151.6&1.78&                        \\ 
ESO440-IG058      &12 06 53.0        &-31 57 08 & 0.023    &11.33   &10.17&99.0&36.3&0.43 &                        \\ 
ESO507-G070       &13 02 51.3        &-23 55 10 & 0.021    &11.44   &10.67&6.8&46.6&0.55  &                       \\ 
IC1623A/B             &01 07 46.3        &-17 30 32 & 0.020    &11.63&10.42&6.0&72.7&0.86 &    {\bf PSN 2011}                       \\ 
IC2545                   &10 06 04.2        &-33 53 04 & 0.034    &11.66&10.58&-0.1&77.7&0.92&                           \\ 
IC2810                  &11 25 47.3        &+14 40 23  & 0.034    &11.60&10.61&1.5&68.0&0.80  &                         \\ 
IC4687/6               &18 13 38.6        &-57 43 36 & 0.017    &11.45&10.43&2.9&47.3&0.56   & {\bf PSN 2010}                        \\ 
IRAS12224-0624  &12 25 02.8        &-06 40 44 & 0.026    &11.30&9.83&2.9&34.0&0.40           &                 \\ 
IRAS14378-3651  &14 40 57.8        &-37 04 25 & 0.068    &12.13&10.40&5.1&233.0&2.74         &                  \\ 
IRAS16399-0937  &16 42 39.2        &-09 43 11 & 0.027    &11.55&10.39&10.0&60.3&0.71         &                   \\ 
IRAS17207-0014  &17 23 21.4        &-00 17 00 & 0.043    &12.42&10.32&-50.&447.5&5.27        &                    \\ 
IRAS18090+0130  &18 11 37.3        &+01 31 40  & 0.029    &11.63&10.54&2.0&71.7&0.84          &                 \\ 
MCG-02-01-051/2 &00 18 51.4        &-10 22 33 & 0.027    &11.41&10.59&3.1&44.4&0.52          &    {\bf 2010 hp}              \\ 
MCG-03-04-014   &01 10 08.5        &-16 51 14 & 0.035    &11.59&10.53&-5.0&66.1&0.78         &                   \\ 
NGC0034             &00 11 06.6        &-12 06 27 & 0.020    &11.43&10.34&-1.0&45.7&0.54     &                      \\ 
NGC0232             &00 42 46.5        &-23 33 31 & 0.020    &11.51&10.71&1.1&55.7&0.66      & {\it 2006et}   \\ 
NGC3110             &10 04 02.7        &-06 28 35 & 0.017    &11.29&10.94&3.3&33.0&0.39      &                           \\ 
NGC5010             &13 12 25.4        &-15 47 45 & 0.021    &10.84&9.79&-1.0&11.8&0.14      &                            \\ 
NGC5331             &13 52 16.6        &+02 06 08  & 0.033    &11.60&10.92&3.0&67.1&0.79      &  \\ 
NGC6240             &16 52 58.6        &+02 24 03  & 0.024    &11.81&10.89&-0.3&108.4&1.28    & {\it 2000bg}, {\bf 2010gp} \\ 
NGC6926             &20 33 04.8        &-02 01 39 & 0.020    &11.25&11.38&5.6&30.4&0.36      &                           \\ 
NGC7130             &21 48 19.6        &-34 57 05 & 0.016    &11.34& 10.75&1.2&37.1&0.44     & {\bf  2010bt}                     \\ 
NGC7592            &23 18 22.2        &-04 24 56 & 0.024    &11.36&10.51&-1.0&38.4&0.45      &                            \\ 
NGC7674             &23 27 56.9        &+08 46 46  & 0.029    &11.37&10.92&1.1&40.0&0.47      & {\it 2000A}, {\bf 2011ee}, {\it 2011hb}  \\ 
\hline
\end{tabular}
\end{centering}
\end{table*}

\begin{table*}
\begin{centering}
\caption{The log of the observations with the epoch of observations (MJD),   the seeing (FWHM in arcsec), and the minimum and maximum magnitude limit for SN detection across the image (cf. Sec.~\ref{artstar}). }\label{log}
\small\addtolength{\tabcolsep}{-2pt}
\begin{tabular}{|cccc||cccc||cccc||cccc|}
\hline
\hline
\multicolumn{4}{|c||}{\bf CGCG011-076}             &\multicolumn{4}{c||}{\bf ESO440-IG058}           &\multicolumn{4}{c||}{\bf IRAS17207-0014}          &\multicolumn{4}{c||}{\bf NGC0232} \\
JD              &Seeing&m$^{lim}_{max}$&m$^{lim}_{min}$&JD              &Seeing&m$^{lim}_{max}$&m$^{lim}_{min}$&JD              &Seeing&m$^{lim}_{max}$&m$^{lim}_{min}$&JD              &Seeing&m$^{lim}_{max}$&m$^{lim}_{min}$\\
\hline
55189.3          &0.41& 19.0        &   13.5      & 54989.0        &0.36 & 19.0         &15.5        &55035.0        &0.39 &19.0         &16.5          &55415.3         &0.34  &19.0         &16.0\\          
55203.2          &0.59& 19.0        &   13.4      &55201.3        &  0.43& 19.0         &15.0        &55056.0        &0.37 &19.0         &16.5          &55704.3         & 1.02  &19.0         &14.0\\
55219.2          &0.65& 19.0        &   13.1      & 55339.0        & 0.56& 19.0         &17.3        &55090.9        &0.56 &19.0         &15.3          &55762.3         &1.05   &19.0         &14.2\\
\cline{1-4}                                                                                                                                             \cline{13-16}                                                        
\multicolumn{4}{|c||}{\bf CGCG043-099}             &55701.9        & 0.49 & 19.0        &18.8        &55342.3        &0.78 &19.0         &14.2         &\multicolumn{4}{c||}{\bf NGC3110}             \\
JD              &Seeing&m$^{lim}_{max}$&m$^{lim}_{min}$&55734.1         & 1.8 & 18.3        &15.5         &55398.1        &1.16 &18.7         &16.7          &JD              &Seeing&m$^{lim}_{max}$&m$^{lim}_{min}$ \\
\cline{1-4}                                       \cline{5-8}                                                                                           \cline{13-16}                                                                                                    
54989.1         &0.49 &19.0          &13.5        &\multicolumn{4}{c||}{\bf ESO507-G070}             &55415.2        &0.49 &19.0        &14.8           &55184.3         &0.47  &19.0         &16.0   \\
55019.1         &0.92 &19.0          &13.0        &JD              &Seeing&m$^{lim}_{max}$&m$^{lim}_{min}$&55446.1        &1.05 &19.0         &16.0         &55202.3         &0.35  &19.0         &16.8  \\ 
                                                  \cline{5-8} 
55238.3         &0.98 &19.0          &13.1        &54989.1         &0.36  &19.0         &16.5        &55675.3        &0.42 &19.0         &18.0           &55291.0         &0.45  &19.0         &15.8   \\
55251.2         &1.16 &19.0          &12.5        &55016.1         &0.53  &  19.0       &16.7        &55739.3        &1.05 &18.5         &17.0          &55338.9         &0.56  &19.0         &16.7   \\
55284.1         &0.42 &19.0          &13.5        &55226.2         &0.36  & 19.0        &16.5        &55775.0        &0.66 &19.0         &17.0          &55662.0         &0.68  &19.0         &15.0   \\
\cline{1-4}
\multicolumn{4}{|c||}{\bf ESO148-IG002}            &55251.3         &0.53  &19.0         &16.5       &55796.1        &1.10 &18.7         &17.3         &55708.0         &0.42  &19.0         &15.5   \\
                                                                                                     \cline{9-12} 
JD              &Seeing&m$^{lim}_{max}$&m$^{lim}_{min}$&55284.0         &0.47  &19.0         &16.7        &\multicolumn{4}{c||}{\bf IRAS18090+0130}          &55189.3         &0.33  &19.0         &16.2    \\
\cline{1-8} \cline{13-16}
54990.3         &0.68  &19.0        &15.0         &\multicolumn{4}{c||}{\bf IC1623A}                 &JD              &Seeing&m$^{lim}_{max}$&m$^{lim}_{min}$          &\multicolumn{4}{c||}{\bf NGC5010}\\
                                                                                                     \cline{9-12}
55028.2         &0.69  &19.0        &15.0         &JD              &Seeing&m$^{lim}_{max}$&m$^{lim}_{min}$&55019.1          &1.19 &19.0         &16.0&JD              &Seeing&m$^{lim}_{max}$&m$^{lim}_{min}$\\
                                                  \cline{5-8}                                                                              \cline{13-16}                                     
55048.4         &0.42  &19.0        &15.0        &55056.3         &0.41  &19.0         &15.5        &55038.1          &0.51 &19.0         &16.7          &55014.9         &0.50  &19.0         &16.0 \\
55098.0         &0.57  &18.7        &15.0        &55071.3         &0.67  &19.0         &15.5        &55056.0          &0.65 &19.0         &16.2          &55035.0         &0.58  &19.0         &14.5 \\
55365.2         &0.56  &18.8        &15.3        &55390.4         &0.54  &19.0         &17.3        &55340.3          &1.00 &18.7         &13.2          &55243.3         &1.02  &18.3         &14.2\\
55375.2         &1.02  &19.0        &14.0        &55415.3         &0.37  &19.0         &16.6        &55340.3          &1.00 &18.7         &13.2          &55284.0         &0.44  &19.0         &17.7 \\
55393.3         &0.99  &19.0        &15.0        &55444.3         &0.51  &19.0         &17.3        &55366.3          &0.58 &19.0         &15.5          &55402.0         &0.99  &19.0         &14.5 \\
55702.3         &0.71  &19.0        &16.8        &55762.3         &0.89  &19.0         &13.5        &55398.1          &1.20 &18.2         &14.8          &55665.1         &0.48  &19.0         &16.3 \\
55719.3         &1.47  &18.2        &15.0        &55796.3         &0.79  &19.0         &15.3        &55415.2          &0.41 &18.8         &16.8          &55710.0         &0.65  &19.0         &16.5 \\
                                                                                                                                                        \cline{13-16}                                                
55743.1         &0.59  &18.8        &16.8        &55812.1         &0.63  &18.8         &14.8        &55672.3          &0.39 &19.0         &15.7          &\multicolumn{4}{c||}{\bf NGC5331}\\
\cline{1-4}                                       \cline{5-8}                                                                                            
\multicolumn{4}{|c||}{\bf ESO239-IG002}           &\multicolumn{4}{c||}{\bf IC2545}                 &55742.2          &0.47 &19.0         &15.3          &JD              &Seeing&m$^{lim}_{max}$& m${lim}_{min}$\\
                                                                                                                                                        \cline{13-16}                                                   
JD              &Seeing&m$^{lim}_{max}$&m$^{lim}_{min}$&JD             &Seeing&m$^{lim}_{max}$&m$^{lim}_{min}$&55796.1          &0.89 &18.5         &15.3          &55016.1         &0.49  &19.0         &14.0\\
\cline{1-4}                                      \cline{5-8}                                        \cline{9-12}
55019.2         &0.72 &19.0         &14.7        &55137.3         &0.95  &19.0         &15.5        &\multicolumn{4}{c||}{\bf MCG-02-01-051}          &55251.3         &0.95  &18.7         &14.8\\
55028.2         &0.57 &19.0         &14.5        &55172.3         &0.48  &19.0         &15.5        &JD              &Seeing&m$^{lim}_{max}$&m$^{lim}_{min}$          &55263.3         &0.48  &19.0         &15.0 \\
                                                                                                    \cline{9-12} 
55056.1         &0.53 &19.0         &14.5        &55662.0         &0.56  &19.0         &16.0        &55037.4         &0.56 &19.0         &18.5             &55665.2         &0.49  &18.8         &16.3 \\
55342.3         &0.63 &19.0         &14.5        &55665.1         &0.57  &19.0         &15.5        &55049.4         &0.47 &19.0         &18.8            &55708.0         &0.42  &17.7         &16.5   \\
                                                  \cline{5-8}                                                                                           \cline{13-16}   
55365.2         &0.69 &19.0         &15.5        &\multicolumn{4}{c||}{\bf IC2810}                  &55068.4         &0.47 &19.0         &18.2           &\multicolumn{4}{c||}{\bf NGC6240}\\
55398.1         &1.16 &18.7         &14.5        &JD             &Seeing&m$^{lim}_{max}$&m$^{lim}_{min}$ &55398.2         &0.92 &19.0         &18.8         &JD              &Seeing&m$^{lim}_{max}$&m$^{lim}_{min}$\\
                                                  \cline{5-8}                                                                                            \cline{13-16}                                                                                                   
55415.2         &0.37 &19.0         &14.7        &55202.3        &0.50  &19.0         &14.0         &55415.3         &0.45 &19.0         &19.0           &55019.1         &0.54  &19.0         &14.7\\
55675.3         &0.50 &19.0         &15.0        &55226.2        &0.51  &19.0         &13.8         &55449.3         &0.47 &18.7         &18.5           &55038.1         &0.53  &19.0         &17.5     \\
55702.2         &0.71 &18.8         &14.5        &55238.3        &0.46  &19.0         &14.8         &55704.3         &1.20 &18.7         &18.3           &55057.0         &0.45  &19.0         &15.8\\
                                                  \cline{5-8}
55719.3         &1.58 &18.2         &14.8        & \multicolumn{4}{c||}{\bf IC4687}                 &55739.2         &1.05 &19.0         &18.7          &55342.3         &0.82  &18.3         &12.3\\
\cline{1-4}                                                                                          \cline{9-12}
\multicolumn{4}{|c||}{\bf ESO244-G012}            &JD             &Seeing&m$^{lim}_{max}$&m$^{lim}_{min}$& \multicolumn{4}{c||}{\bf MCG-03-04-014}            &55398.0         &1.00  &17.7         &12.3\\
                                                  \cline{5-8}
JD              &Seeing&m$^{lim}_{max}$&m$^{lim}_{min}$&54989.1         &0.37  &19.0         &16.0       &JD              &Seeing&m$^{lim}_{max}$&m$^{lim}_{min}$           &55416.1         &0.36  &18.2         &11.0\\
\cline{1-4}                                                                                           \cline{9-12}                                                             
55028.3         &1.11 &19.0         &13.0        &55056.0         & 0.65  &19.0         &15.8       &55056.3         &0.48 &19.0         &17.0             &55455.9         &0.55  &18.7         &13.0\\
55056.4         &0.39 &19.0         &14.0        &55337.1         &0.55  &19.0         &16.2        &55106.2         &0.74 &19.0         &16.2 &55771.1         &0.47  &19.0         &15.0\\
                                                                                                                                                     
55071.4         &0.63 &19.0         &13.5        &55366.3         &0.58  &19.0         &16.2       &55443.3         &0.84 &19.0         &14.8             &55785.1         &1.03  &19.0         &14.7\\
55375.3         &1.16 &18.0         &13.7        &55375.0         &1.21  &19.0         &16.2       &55762.3         &1.05 &19.0         &15.8            &55796.0         &1.26  &19.0         &13.5\\
                                                                                                                                                         \cline{13-16}                                            
55393.3         &1.20 &18.2         &14.0        &55393.2         &1.21  &19.0         &15.3       &55796.3         &0.63 &19.0         &14.8            &\multicolumn{4}{c||}{\bf NGC6926}\\
55402.2         &0.82 &18.8         &13.5        &55663.3         &0.66  &19.0         &16.2       &55812.1         &0.53 &19.0         &15.0            &JD              &Seeing&m$^{lim}_{max}$&m$^{lim}_{min}$\\
                                                  \cline{5-8}                                                                                            \cline{13-16}                                                     
55705.3         &1.26 &18.5         &15.5        &  \multicolumn{4}{c||}{\bf IRAS12224-0624}       &55402.2         &1.16 &19.0         &14.5            &55068.1         &0.44  &19.0         &14.8\\
55757.3         &0.84 &19.0         &16.2        &JD             &Seeing&m$^{lim}_{max}$&m$^{lim}_{min}$&55415.4         &0.39 &19.0         &14.8            &55366.3         &0.43  &19.0         &15.0\\
                                                   \cline{5-8}                                        \cline{9-12} 
\cline{1-4}                                                
\multicolumn{4}{|c||}{\bf ESO264-G036}            &54989.0         &0.51&19.0         &16.5        & \multicolumn{4}{c||}{\bf NGC0034}                   &55398.1         &1.26  &18.5         &14.0\\
JD              &Seeing&m$^{lim}_{max}$&m$^{lim}_{min}$&55019.0         &0.84&19.0         &16.0        &JD              &Seeing&m$^{lim}_{max}$&m$^{lim}_{min}$            &55415.2         &0.34  &19.0         &16.3\\
\cline{1-4}                                                                                           \cline{9-12}
55142.3         &0.98 &18.0         &14.0        &55226.2         &0.47&19.0         &16.0         &55049.3         &0.47 &18.5         &14.3            &55672.3         &0.39  &19.0         &16.7\\
                                                   \cline{5-8}                                        
55172.3         &0.63 &18.3         &14.5        &  \multicolumn{4}{c||}{\bf IRAS14378-3651}       &55068.3         &0.53 &18.8         &14.0                  &55701.3         &0.74  &18.7         &15.5\\
55184.3         &0.45 &18.7         &14.0         &JD             &Seeing&m$^{lim}_{max}$&m$^{lim}_{min}$&55398.2         &1.37 &18.2        &12.7               & 55743.1        &0.45  &19.0         &17.7\\
                                                   \cline{5-8}                                                                               \cline{13-16}   
55339.0         &0.59 &18.3         &13.5        &54990.1         &0.53&19.0         &13.5        &55449.3         &0.40 &19.0         &13.5             &\multicolumn{4}{c||}{\bf NGC7130}\\
55665.1         &0.55 &19.0         &16.8        &55072.9         &0.36&19.0         &14.5        &55704.4         &0.79 &19.0         &14.0             &JD              &Seeing&m$^{lim}_{max}$&m$^{lim}_{min}$\\
                                                   \cline{5-8}                                                                                           \cline{13-16}                                        
55701.9         &0.50 &19.0         &17.5        &  \multicolumn{4}{c||}{\bf IRAS16399-0937}      &55739.2         &1.00 &19.0         &15.3             &55038.1         &0.47  &19.0         &16.0 \\
55734.0         &1.80 &18.8         &16.7        &JD             &Seeing&m$^{lim}_{max}$&m$^{lim}_{min}$&55771.3         &0.59 &18.8         &14.2            &55341.3         &0.76  &19.0         &14.7\\
\cline{1-4}                                        \cline{5-8}                                     \cline{9-12}  
\multicolumn{4}{|c||}{\bf ESO286-IG019}           &55014.1         &0.49 &19.0         &16.0      & \multicolumn{4}{c||}{\bf NGC7674}            &55366.3         &0.44  &19.0         &16.3\\
                                                                                                   
JD              &Seeing&m$^{lim}_{max}$&m$^{lim}_{min}$& 55035.0         &0.42&19.0         &16.7      & JD              &Seeing&m$^{lim}_{max}$&m $^{lim}_{min}$                    &55382.2         &0.57  &19.0         &16.7\\
\cline{1-4}                                                                                         \cline{9-12}    
54988.4         &0.34  & 19.0       &15.0         &55057.0         &0.38 &19.0         &16.3      &55068.1         &0.47 &19.0         &14.5             &55675.3         &0.40  &18.5         &13.7\\
55019.1         &0.62  & 19.0       &15.5         &55091.0         &0.59 &19.0         &16.8      &55398.1         &1.24 &18.0         &12.7            &55719.2         &1.37  &18.2         &15.3\\
55038.1         &0.45 & 19.0       &15.0         &55342.3         &0.80 &19.0         &15.7       &55415.3         &0.42 &19.0         &13.8           &55735.2         &1.0   &19.0         &16.0\\
                                                                                                                                                         \cline{13-16}                                          
55056.1         &0.45 & 19.0       &15.5         &55398.0         &1.20 &18.7         &15.7       &55446.1         &0.84 &19.0         &14.0           &\multicolumn{4}{c||}{\bf NGC7592}\\
                                                                                                                                                                 
55341.4         &0.74 & 19.0       &14.5         &55416.1         &0.40  &19.0         &16.5      &55739.3         & 0.68 &19.0         &15.7          &JD              &Seeing&m$^{lim}_{max}$&m$^{lim}_{min}$\\
                                                                                                                                                         \cline{13-16}                                          
55365.1         &0.49 & 19.0       &15.2         &55446.0         &1.00  &19.0         &16.3      & 55796.3        &0.74 &19.0         &15.3                                                &55398.2         &1.0  &17.5         &13.8\\  
                                                                                                    \cline{9-12}
55382.2         &0.63 & 19.0       &15.0         &55665.2         &0.60  &19.0         &17.0      & \multicolumn{4}{c||}{\bf NGC0232}                     &55415.2         &0.37  &18.7         &14.2  \\
55398.1         &0.84 & 18.7       &15.0         &55710.0         &0.63 &18.7         &16.3       & JD              &Seeing&m$^{lim}_{max}$&m$^{lim}_{min}$            &55444.2         &0.40  &19.0         &15.0\\
                                                                                                     \cline{9-12}
55700.2         &0.65 & 19.0       &16.8         &55771.1         &0.37 &18.3         &15.7       &55090.1         &0.37 &19.0         &15.2             &55704.3         &0.95  &19.0         &13.5\\
55719.2         &1.34 & 18.5       &15.2         &55785.0         &0.89  &18.8         &16.2      &55382.3         &0.69 &19.0         &14.7             &55771.3         &0.47  &19.0         &15.5\\
55735.2         &1.60 & 19.0       &14.5         &55795.0         &0.87  &18.9         &16.2      &55402.2         &0.95 &19.0         &16.0             &55796.3         &0.65  &19.0         &13.5\\
\hline
\end{tabular}
\end{centering}
\end{table*}
\subsection{Data reduction and analysis}\label{dataredu}

For  data reduction and  mining of  the HAWK-I mosaic images we developed a custom pipeline that integrates different, publicly available, recipes and tools in a {\em Python} environment.

The pipeline consists of  four sections: 

\begin{enumerate}
\item pre-reduction,  astrometric calibration and production of the stacked mosaic image. For these steps we use the ESO HAWK-I pipeline recipes in  {\em EsoRex}, the ESO Recipe Execution Tool\footnote{http://www.eso.org/sci/software/cpl/esorex.html};
\item subtraction of images taken at different epochs using {\em ISIS} \citep{alard:2000tn} for the PSF matching;
\item search for transient candidates in the difference image using {\em Sextractor} \citep{bertin:1996fk}.
The candidates were ranked based on their {\em Sextractor} measured parameters and submitted to the operator for visual inspection and validation;
\item estimate of the detection efficiency through artificial star experiments performed for each of the search images (details in Sect.\ref{artstar}).
\end{enumerate}

The raw images were retrieved from the ESO archive as soon as they became available, and immediately reduced to allow for activation of follow-up spectroscopy of  transient candidates.

For the pre-reduction, we followed the reduction cascade described in the HAWK-I pipeline manual\footnote{ftp://ftp.eso.org/pub/dfs/pipelines/hawki/hawki-pipeline-manual-1.8.pdf}  including dark subtraction, flat field and illumination corrections, background subtraction, distortion correction,
astrometric offset refinement, combination of the different exposures and stitch of the 4 detectors in a single mosaic image. Actually, it turned out that the ESO pipeline recipes for background subtraction and offset refinement do not provide satisfactory results for our images. The main reason is  the extended size of our  sources and the consequent large dithering we had adopted. To address this issue we implemented custom recipes for the two afore mentioned reduction steps.

The most critical step of the data reduction is the image subtraction, in particular in the proximity of the nuclear regions of the galaxies. First of all we need to choose a proper reference image, usually the image with the best seeing obtained at least three month before (or in some case after) the image to be searched. We also need to choose the proper parameters for the image difference procedure (see \citealt{melinder:2012ly} for an extensive discussion). An additional problems arises because   in the distributed version of {\em ISIS}, the program automatically selects the reference sources for the computation of the convolution kernel. Owing to the small number of sources in our extragalactic fields, the reference source list in general includes the bright galaxy nucleus which, being very bright, has a significant weight in the determination of the kernel. This may cause some problems because if at
one epoch a SN occurs very close to the galaxy nucleus it can be included in the convolution kernel and effectively cancelled in the difference image.  We therefore modified the {\em ISIS} selection procedure to allow for exclusion of specific sources,  in particular the galaxy nuclei, from the reference list.

\begin{figure}
\centering
\includegraphics[width=.5\textwidth]{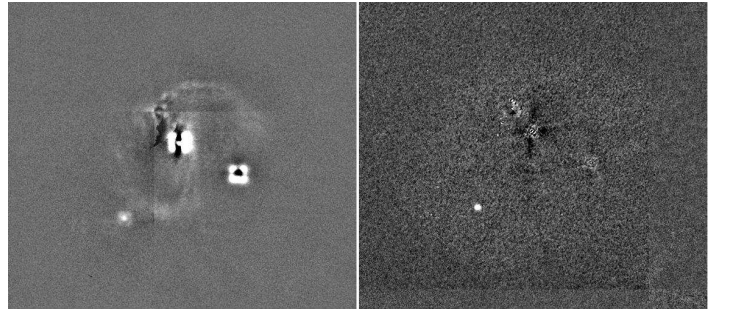}
\caption[]{\emph{Left panel}: example of poor subtraction of  images  of NGC~7130 with large seeing differences ($FWHM =1.5\arcsec$, with $FWHM=0.4\arcsec'$ for the reference).
 \emph{Right panel}:  optimal subtraction for two images with similar, good seeing ($FWHM=0.4\arcsec$). In both panels the source in the lower left quadrant is SN~2010bt (cf. Fig.~\ref{maps}).  The FOV in both panel is about 2'$\times$2'.}  \label{dseeing}
\end{figure}

Despite the efforts  in many cases the difference image shows significant spurious residuals 
in correspondence to the galaxy nuclear regions. The problem is most severe in case of images with poor seeing ($FWHM> 1 \arcsec$)  and/or reduced transparency.
This is illustrated in Fig. \ref{dseeing} where we show two examples of image difference one for a search image with poor seeing ($FWHM=1.5\arcsec$, left panel)  and the other for a case with excellent seeing ($FWHM=0.4\arcsec$). In both cases, the reference image was the same and had excellent seeing  ($FWHM=0.4\arcsec$).\\
 False detections due to residuals of the image subtraction were largely removed by the requirement that the candidate had to be visible at least in two consecutive epochs.

\subsection{Supernova discoveries and characterization}\label{sndisc}

\begin{figure*}
\centering
\subfloat[SN 2010bt]{\includegraphics[width=.4\textwidth]{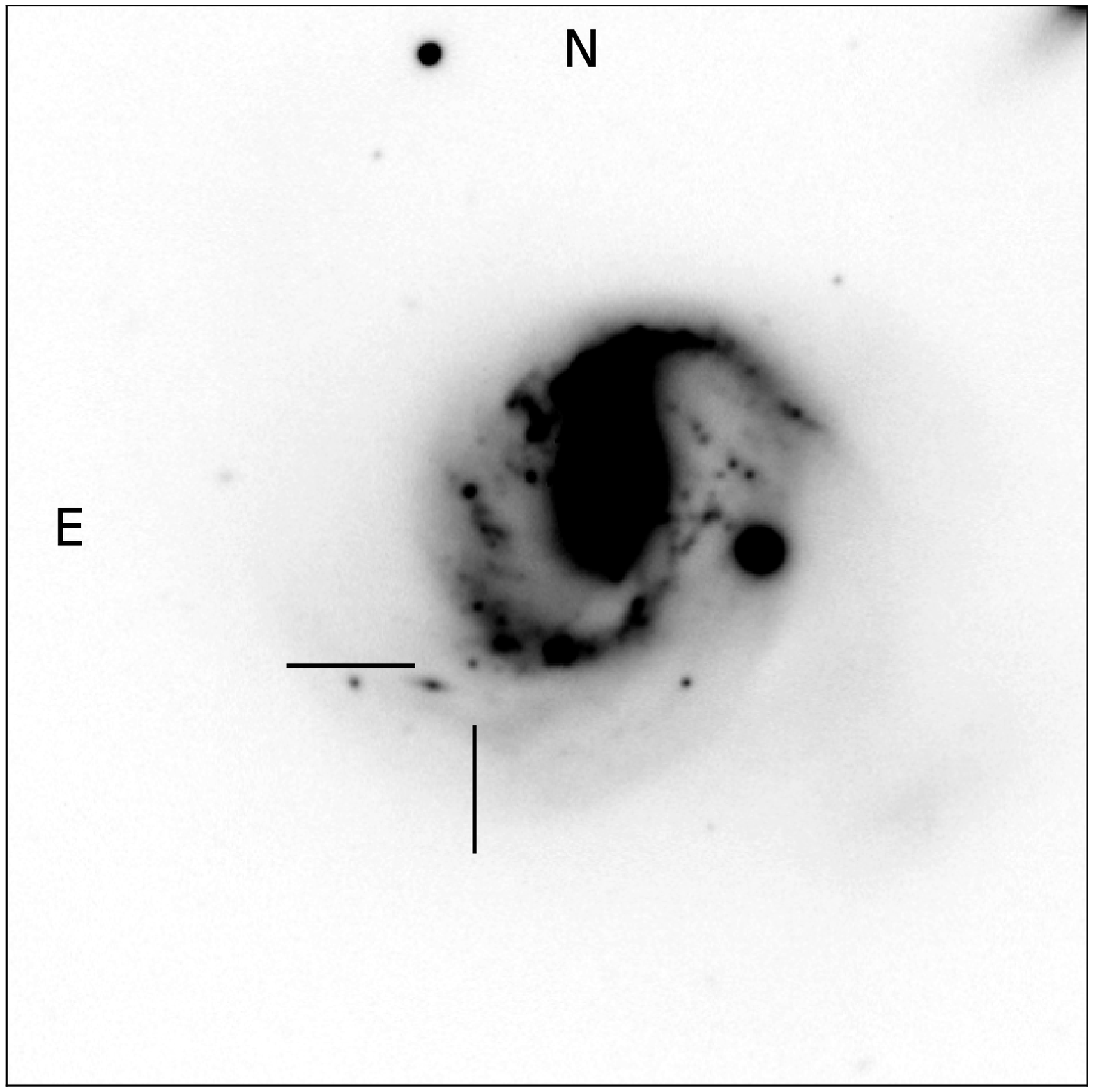}}\qquad
\subfloat[SN 2010gp]{\includegraphics[width=.4\textwidth]{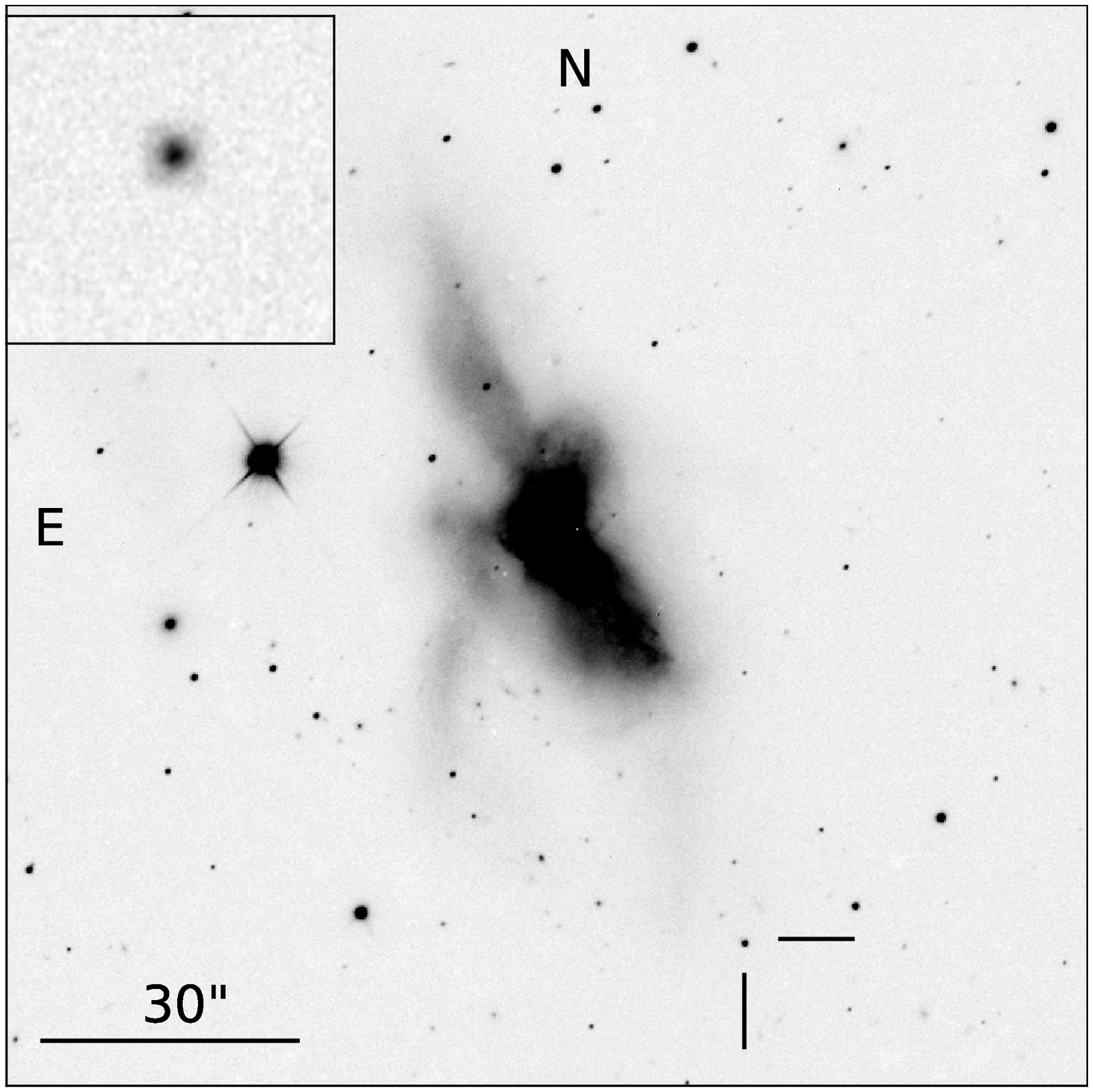}}\\
\subfloat[SN 2010hp]{\includegraphics[width=.4\textwidth]{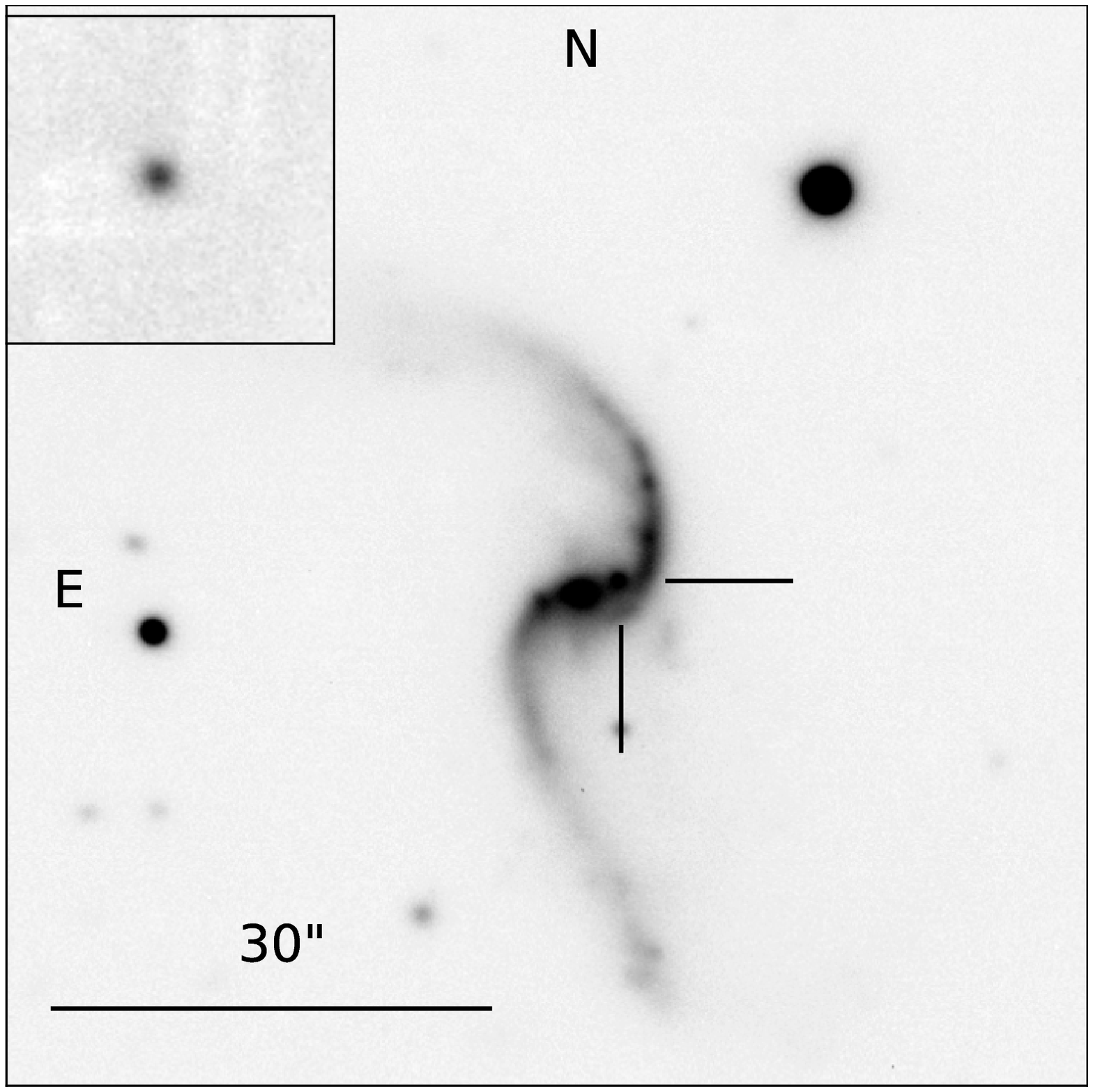}}\qquad
\subfloat[PSN2010 in IC 4687]{\includegraphics[width=.4\textwidth]{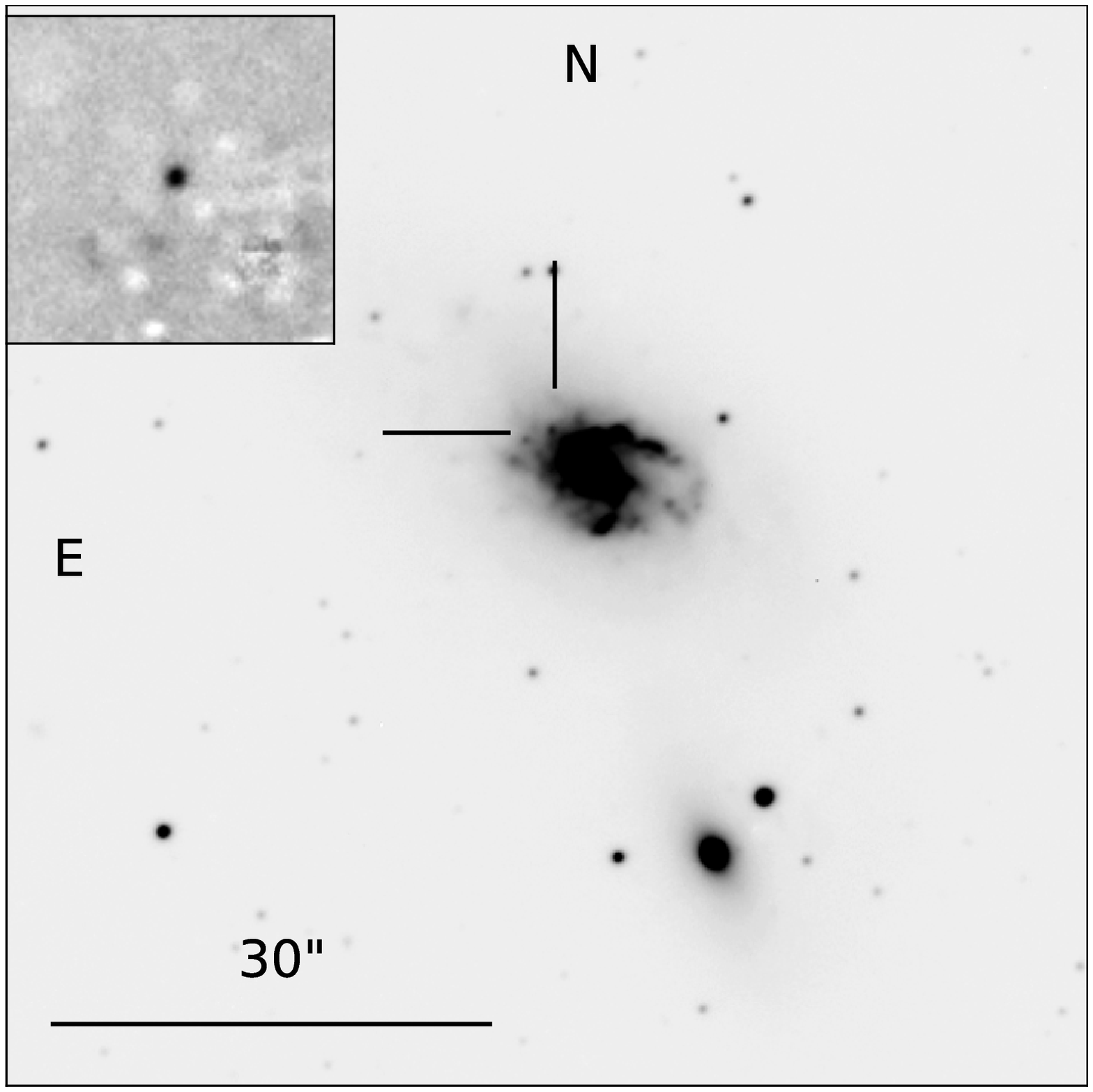}}\\\subfloat[SN 2011ee]{\includegraphics[width=.4\textwidth]{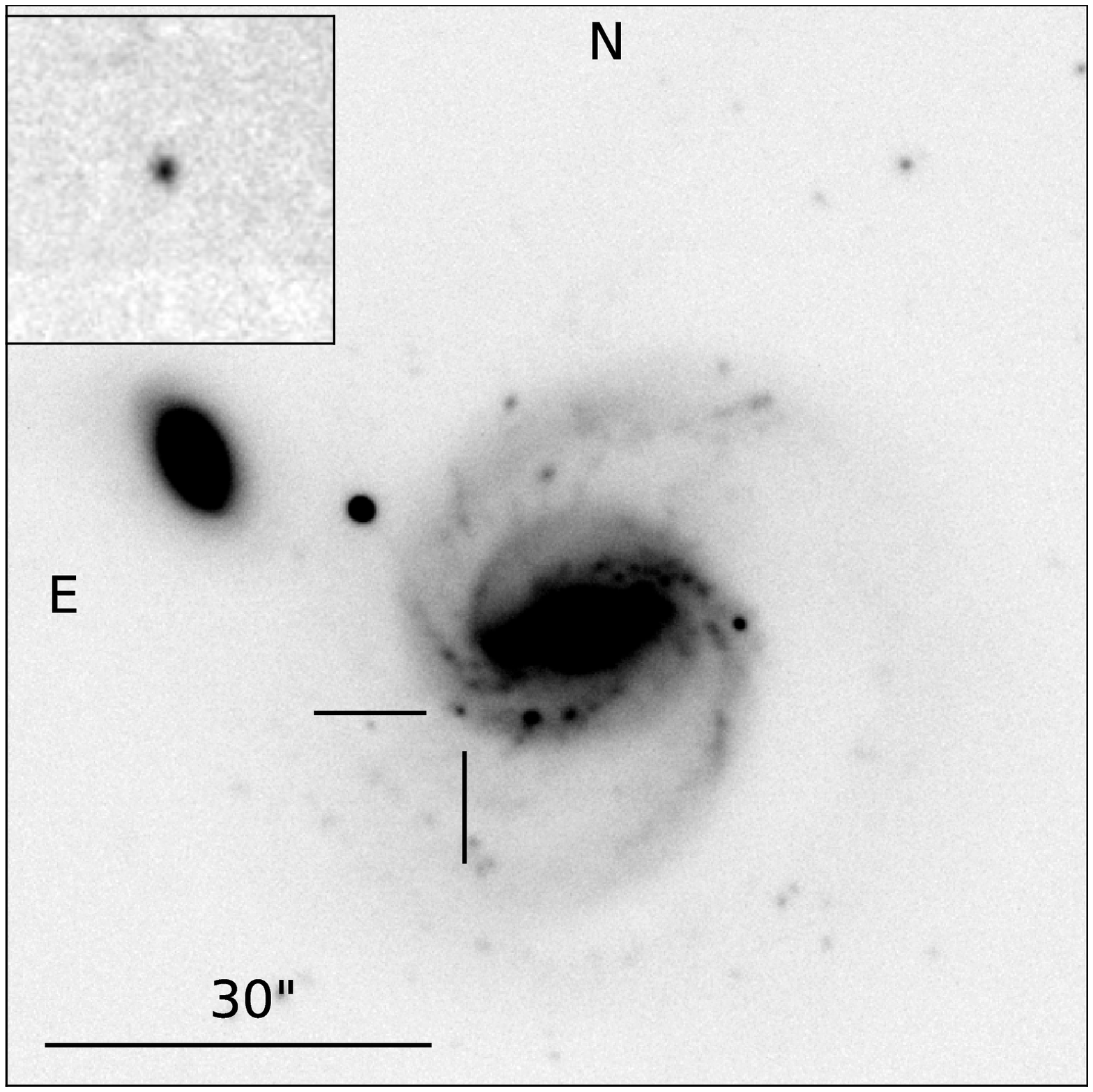}}\qquad
\subfloat[PSN 2011 in IC1623A]{\includegraphics[width=.4\textwidth]{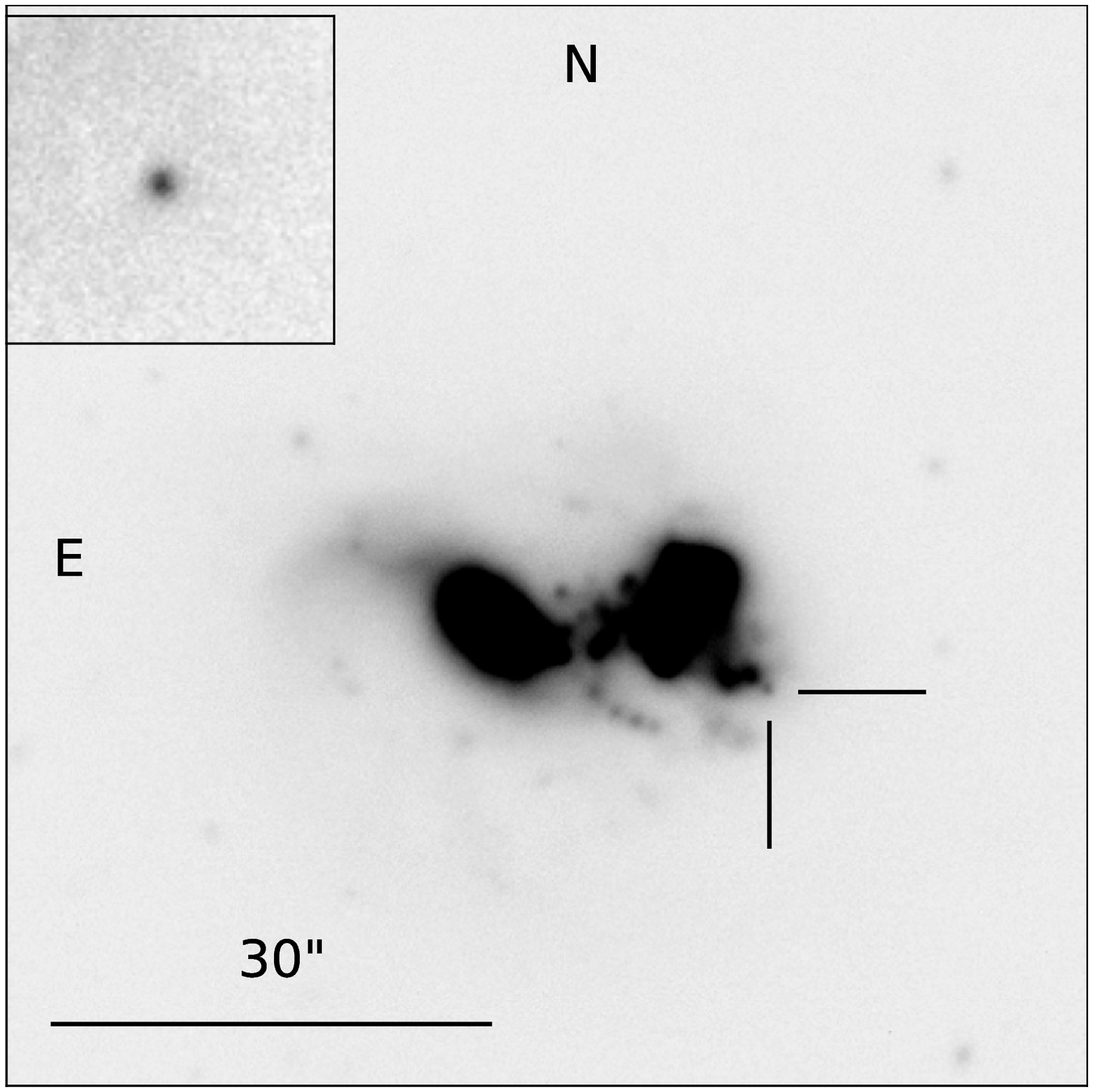}}
\caption{K-band finding charts for the SNe of our list. The inserts show the transients as they appear in the difference image. }\label{maps}
\end{figure*}

During our monitoring campaign  6 transients were detected in at least two consecutive epochs separated by at least one month (finding charts are in Fig.~\ref{maps}). Four of them were spectroscopically confirmed as SNe (three SN-CC and one SN~Ia) and we will argue in the following that also the other two transients, labeled as probable SN (PSN), are  likely SN~CC (Tab.~\ref{snelist}). SNe 2010bt  and 2010gp  were discovered and announced before our detection by optical  searches but have been independently re-discovered by us. 

The objects are listed in Tab.~\ref{snelist} along with the  host galaxy name, distance modulus (computed from the galactocentric redshift and the adopted cosmology), SN   coordinates,  offsets from the galaxy nucleus and  projected linear distances  from the galaxy nucleus.

\begin{table*}
\begin{centering}
\caption{Information for the detected SNe. }\label{snelist}
\begin{tabular}{l|cccccc}
\hline
                     &  SN2010bt  & SN2010gp & SN2010hp       & SN2011ee & PSN2010 &  PSN2011 \\
\hline
Host galaxy&  NGC7130  & NGC6240  &MCG-02-01-52  & NGC7674  & IC 4687 & IC 1623A    \\  
Distance modulus& 34.08      & 35.07       &  35.24            & 35.39           &34.21           &34.54\\
R.A. (SN)      & 21:48:20.22 & 16:52:57.39  & 0:18:50.01   & 23:27:57.34     &18:13:40.213     & 01:07:46.229\\
Dec. (SN)   & 34:57:16.5  &2:23:16.4       &-10:21:40.6    & +08:46:38.10   &-57:43:28.00    &-17:30:29.48\\
Offset [\arcsec] & 9E~14S  & 22W~47S  &2.6W~2.7N      &9.3E~6.3S       &2.6E~2.8N       &4E~7S  \\
$r$  [Kpc]          & 5.3        & 25.9           &2.1            & 8.9            &6.5             &3.2  \\
SN type       &   IIn       &   Ia              &IIP               &Ic              &IIP$^\#$              &Ic$^\#$          \\
R max [mag] & $15.9 \pm0.1$   & $16.0\pm0.1$   &$17.6\pm0.2$     & $17.3\pm0.5$       &                & $16.9\pm0.5$\\
K max  [mag] & $16.1\pm0.2$   & $16.9\pm0.1$    &$17.2\pm0.1$      & $18.0\pm0.5$      & $17.9\pm0.2$      &$17.0\pm0.3$\\
$AB_G$ [mag]&0.44        &0.33            &0.16           &0.25            &0.43            &0.07 \\
$AB_H$ [mag]&  $1.7 \pm 0.5 $ &$0.2\pm0.1$    &$0.5\pm 0.3$ &$\sim 0.0$        &   $0-8$   &  $0.5\pm0.5$ \\
abs $M_R$&-19.2       &-19.2          &-17.9          &-18.1           &                    &-17.9  \\
abs $M_K$&-18.1       &-18.2           &-18.0          &-17.7           &-16.3  / -17.3         & -17.5 \\
MJD max R &$55303 \pm 5$& $55405 \pm 3$       &$55399 \pm 5$      &$55760 \pm 5$       &$55270 \pm 20$       & $55725 \pm 10$  \\
Refs  & CBET 2250, 2252& CBET 2388, 2390&CBET 2246, 2249 & CBET 2773       &                 &     \\
\hline
\end{tabular}

\# Photometric classification.
\end{centering}
\end{table*}

For all  transients K-band magnitudes were measured through  aperture photometry on the difference images and calibrated with respect to  2MASS stars in the field. 
Upper limits measured on  pre-discovery images were also estimated. For all transients, but  PSN2010 in  IC4687, we obtained some follow-up imaging in the optical or near-infrared domains.  These  observations were reduced using standard procedures in {\em IRAF}. When a reference image was not available, the SN magnitude was measured using the PSF fitting technique. Optical band magnitudes were calibrated with respect to Landolt's standard fields.  Our photometry for the  six transients is reported in Tab.\ref{klc}.  For the two transients with no spectroscopic confirmation, the photometry will be used to assess their nature. 

\begin{table*}
\centering
\caption{Transient photometry. Estimated errors are given in parentheses. }\label{klc}
\begin{tabular}{ccccccccc}
\hline
    MJD      &B             &V             &R             & I            & J         &H         &K      & Instr.         \\
\hline
\multicolumn{9}{c}{\bf SN 2010bt}           \\
 55341.4 &               &              &              &              &           &          & 16.9~ (.1) & H \\
 55366.4 &               &              &              &              &           &          & 18.08 (.1) & H \\
 55382.3 &               &              &              &              &           &          & 18.68 (.1) & H \\
\hline
 \multicolumn{9}{c}{\bf SN 2010gp}           \\
55342.4 &               &             &               &             &            &         &  $>18.5$    &  H\\
55398.1 &               &              &              &              &           &         & 17.4~ (.1) & H \\
55416.1 &               &              &              &              &           &         & 16.85 (.1) & H \\
55457.0 &               &              &              &              &           &         & 17.9~ (.1) & H \\
55455.0 &19.72 (.05) &18.43 (.03)  &18.11 (.03) &18.15 (.03) &           &         &    & E\\
\hline
\multicolumn{9}{c}{\bf 2010hp}\\ 
55398.3     &               &              &              &              &           &         & 17.25 (.05) & H  \\
55415.4     &               &              &              &              &           &         & 17.17 (.05) & H  \\
 55449.4     &               &              &              &              &           &         & 17.20 (.10) & H  \\
 55454.3   &19.61 (.03)&18.70 (.03)&18.28 (.03)&17.89 (.03)&           &         & &E\\
 55456.1   &              &              &           &                  &17.24 (.1)&16.93 (.1)&17.17 (.1)& S\\
 55498.1    &              &              &           &                  &18.48 (.1)&18.51 (.1) &17.99 (.1)&S\\
 55499.2  &$>$19.1   &19.61 (.03)&18.83 (.03)&18.83 (.03) &           &         &  &E\\
 55563.1   &$>$20.8    &$>$20.0    &$>$18.5    &19.50 (.05) &           &         &   & E\\
 55737.3   &              &              &$>$20.4    &$>$20.1    &           &         & & E\\
\hline
\multicolumn{9}{c}{\bf 2011ee} \\ 
 55739.3 &               &              &              &              &           &         & 18.6 (.2) & H \\
 55752.3   &               &              &18.50(.1)&                &           &         &  &R\\
 55771.4  &               &              &              &              &           &         & 18.4 (.1) & H  \\
55775.4  &               &              &              &              &           &          &18.5 (.1) & H \\
55796.3  &               &              &              &              &           &          &19.2 (.1) & H \\
55808.3   &$>$19.72(0.05)&18.43(0.03)&18.11(0.03)&18.15(0.03) &        &         & &E\\\hline
\multicolumn{9}{c}{\bf PSN 2010} \\
   55056.1  &               &              &              &              &           &         & $>$19.3 & H  \\
   55336.2 &             &                &              &              &           &          &17.98 (.10) & H\\ 
  55337. 2 &               &              &              &              &           &         & 17.99 (.10) & H  \\
  55338. 2 &               &              &              &              &           &         & 17.91 (.10) & H  \\
   55366.3  &               &              &              &              &           &         & 17.85 (.10)& H \\
   55375.1  &               &              &              &              &           &         & 18.42 (.3) &H \\
   55393.3  &               &              &              &              &           &         & 18.77 (.3) &H \\
\hline
\multicolumn{9}{c}{\bf PSN2011} \\
 55762.4  &               &              &              &              &           &         & 18.15 (.2) & H  \\
  55774.7  &               &$>$19.30    &19.50 (.2)&19.0 (.2)  &           &         & &D\\
  55796.4  &               &              &              &              &           &         & 18.72 (.2) & H  \\
   55808.4   &$>$21.30    &$>$20.0    &20.10 (.3) &19.50(.3)&           &         &  &E\\
55812.2  &               &              &              &              &           &         & 19.43 (.2) &H \\
\hline
\end{tabular}

H = HAWK-I@VLT, E = EFOSC2@NTT, S = SOFI@NTT, L = RATCam@Liverpool, D = Dolores@TNG
\end{table*}

Spectroscopic observations were obtained for four candidates: epoch, spectral range and instruments are reported in  Tab.\ref{spectra}. Data were reduced using standard procedure in {\em IRAF} but for  the X-Shooter spectra which were reduced using version 1.0.0 of the ESO X-shooter pipeline
\citep{goldoni:2006gr} with the calibration frames (biases, darks, arc lamps, and flat fields)
taken during daytime.
The extracted spectra, after wavelength and flux calibration, were compared with a library of template  spectra using the  {\em GELATO} SN spectra comparison tool  \citep[https://gelato.tng.iac.es/,][]{harutyunyan:2008ez}.
The best fit template SN, the SN type and phase are reported in Tab.\ref{spectra}.
 Spectroscopic classification for PSN2010 in IC4687 was attempted, but the observed spectrum resulted too noisy for a safe classification. 
 The table includes the result of the spectroscopic observations of SN 2010gp from \cite{folatelli:2010wk}.

\begin{table*}
\begin{center}
\caption{Log of spectroscopic observations.}\label{spectra}
\begin{tabular}{lccccccc}
\hline
SN                    &    MJD        & range(nm) & res. (nm)  & Instrument            & best fit &  type & Phase\\
\hline
2010bt             &    55304.4    & 350-1000  & 1.4     & EFOSC/NTT         & 2005gj   & IIn    & max    \\
2010hp           &    55454.2     & 350-1000  & 1.4    & EFOSC/NTT          & 1999em & Ia     &+60d   \\
2010 IC4687 &    55352.2 &   340-1000   & 0.2     & XSHOOTER/VLT  &\multicolumn{3}{c}{S/N too low}\\
2011ee      &   55824.6 &              & 2.5 &       OSIRIS/GTC &  2007gr &  Ic & +60\\
2011ee           &   55759.3     & 340-1000   & 0.2      & XSHOOTER/VLT  & 1994I  & Ic     & max \\
2010gp           &  55401.0 &   354-886 & \multicolumn{2}{c}{\cite{folatelli:2010wk}}&  2002bo & Ia & just before max\\
\hline
\end{tabular}
\end{center}
\end{table*}

We have used the available photometry and spectroscopy to put some constraints to the amount of extinction suffered by the SNe.  Hereafter we will describe in some details the sparse information available for each transient.

\begin{description}

\item[{\bf SN 2010bt}]   was discovered on 2010 April 17.10 UT by \cite{monard:2010gp}.  A spectrum taken on April 18.39 UT \citep{turatto:2010aw} shows strong resemblance to several type-IIn SNe, in particular SN~1996L \citep{benetti:1999kx} shortly after explosion (Fig.~\ref{spec}).  A broad H$_{\alpha}$ component is present indicating an expansion velocity of about 3500 ${\rm km}\,{\rm s}^{-1}$ (half width at zero intensity).  SN 2010bt was independently {\bf re-}discovered by us on May 25 and was observed in other 2 epochs. The object was not visible on a  HAWK-I image taken in 2009 July 26 (limit $K=19.0$ mag).  While the analysis of the  light curve and spectral evolution of this SN will be presented elsewhere \citep{elias-rosa:2013vn}, a preliminary analysis shows that to match the color of SN 2010bt to that of the type IIn SN 1998S requires a significant amount of extinction ($A_B = 1.7 \pm 0.5$ mag). This extinction improves also the matching of the spectra  of SN 2010bt to  SN 1996L.

\begin{figure}
\centering
\subfloat[The spectrum of 2010bt, dereddened by $A_B = 1.7$ mag (see text) is compared with that of the type IIn SN 1996L.]{\includegraphics[width=.4\textwidth]{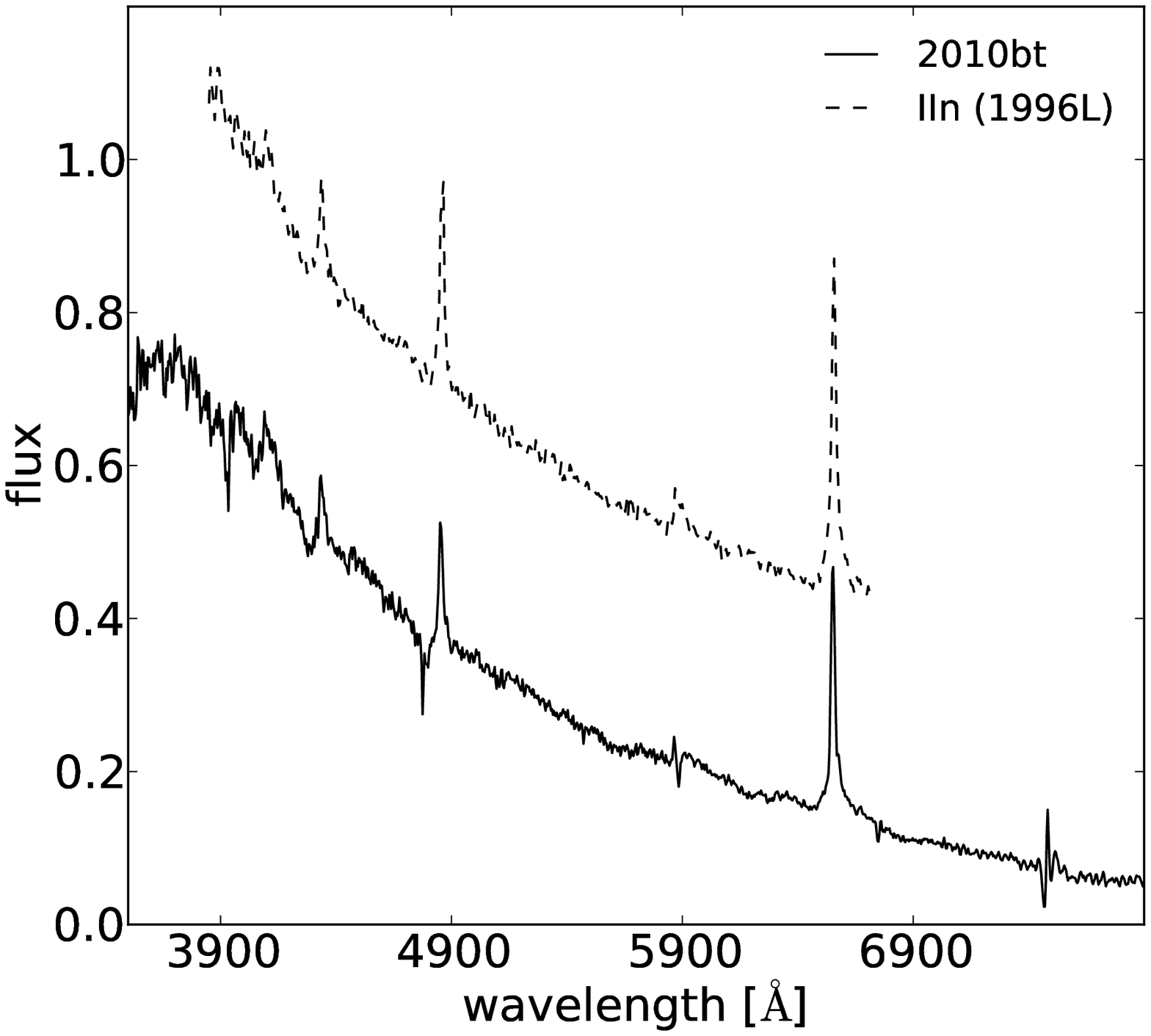}}\\
\subfloat[The spectrum of 2010hp, dereddened by $A_B = 0,5$ mag (see text) is compared with that of the type IIP SN~1999em.]{\includegraphics[width=.4\textwidth]{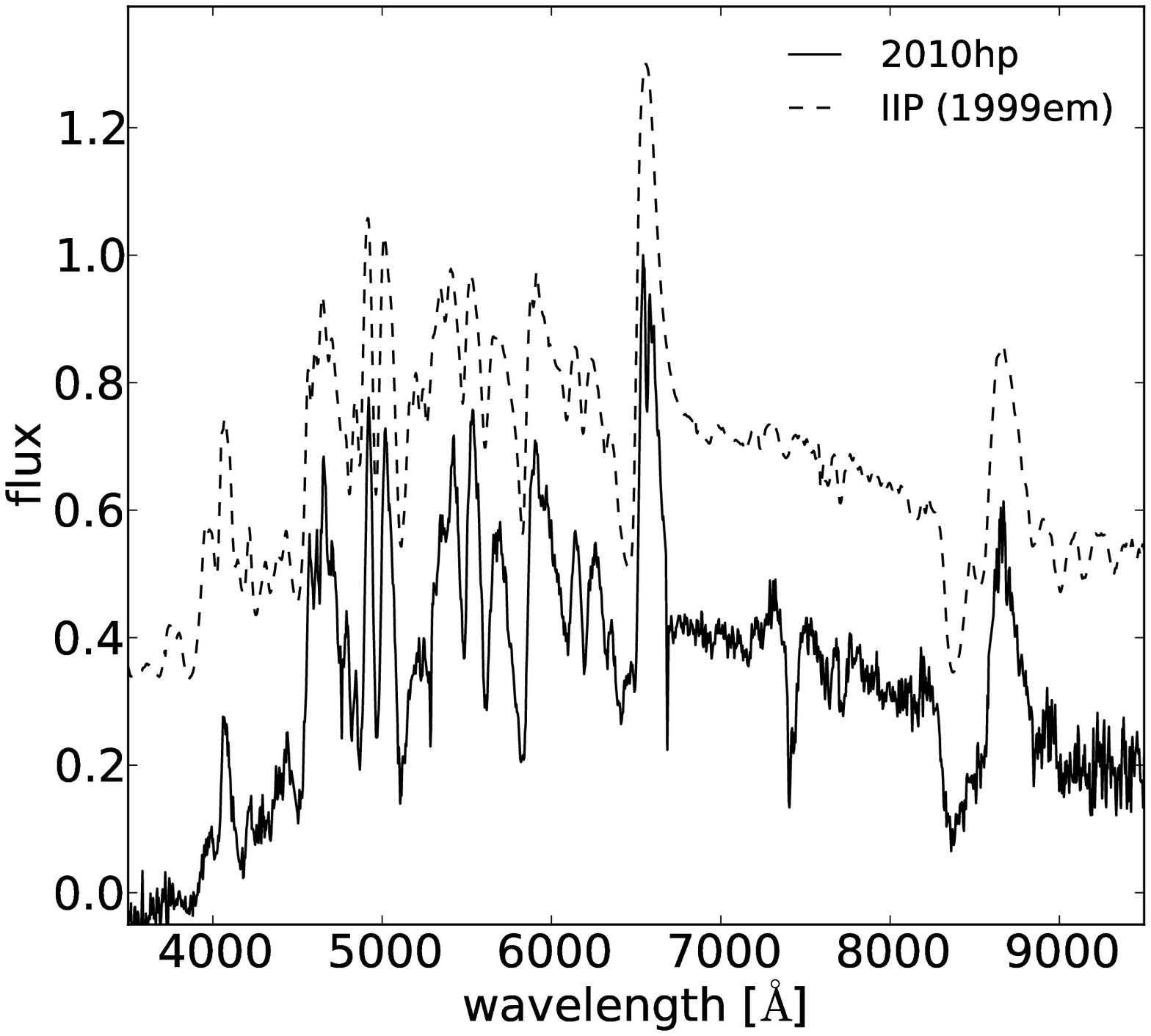}}\\
\subfloat[The spectrum of 2011ee is compared to that of the SN~Ic 2007gr  at the maximum (top panel) and 60 days after the maximum.]{\includegraphics[width=.40\textwidth]{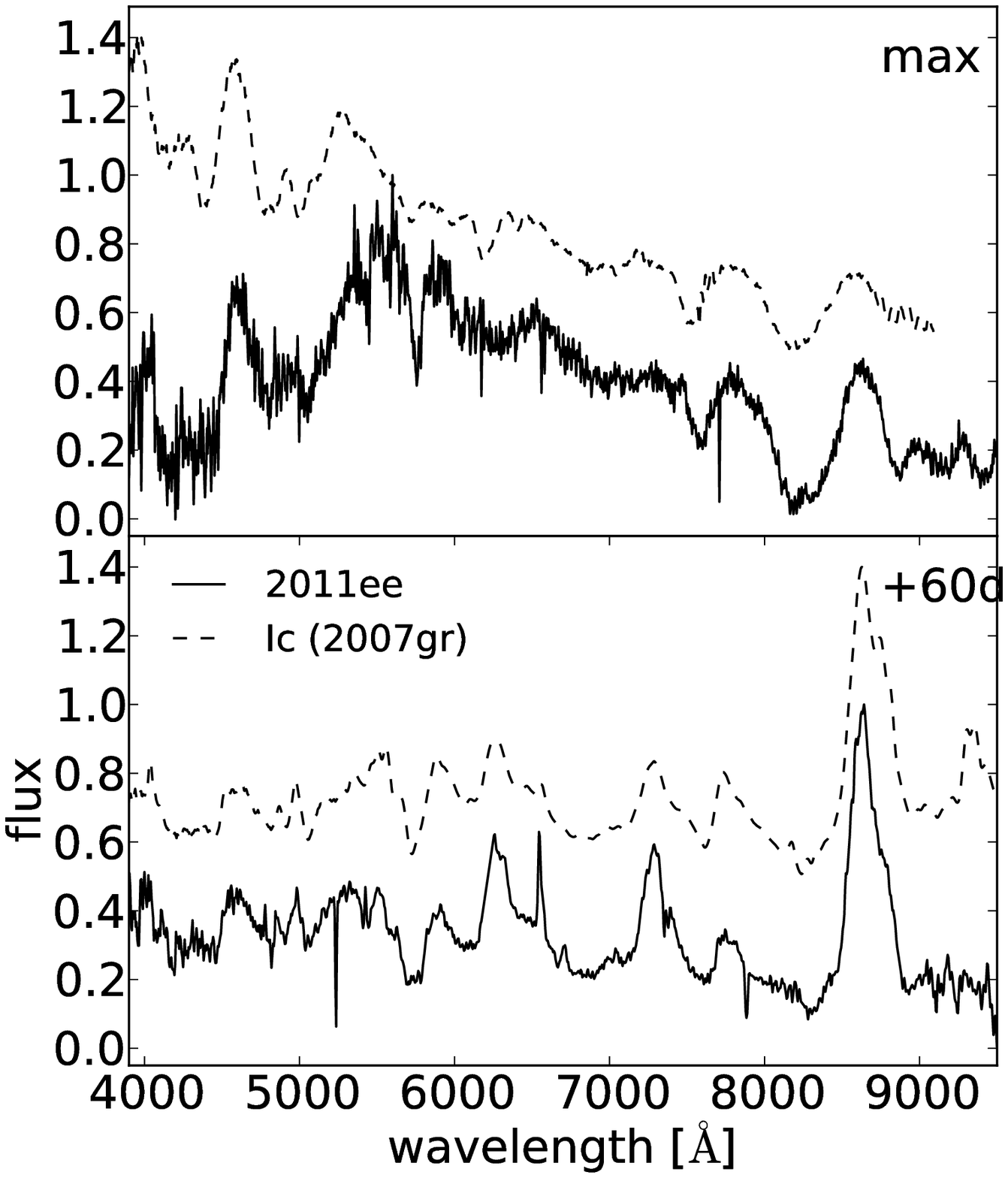}}
\caption{Spectra of SN 2010bt, 2010hp, and 2011ee are shown along with  the best fitting templates.}\label{spec}
\end{figure}

\item[{\bf SN 2010gp}] was discovered on 2010 July 14.10 UT by \cite{maza:2010bl} with  the 0.41-m PROMPT1 telescope located at Cerro Tololo. \cite{folatelli:2010wk} reported the spectroscopic classification as a type-Ia SN around  maximum light and with high expansion velocity of the ejecta. 
SN 2010gp was {\bf re-}discovered independently by us on July 21 and was observed in other 2 epochs. The object was not visible on a HAWK-I image taken on 2010 May 26 (limit $K=18.5$ mag).
The available colors compared to that of standard SN Ia suggest a small reddening inside the host galaxy, $A_B = 0.2$ mag.

\item[{\bf SN 2010hp}] was discovered on a HAWK-I image taken on 2010 July 21.3 UT \citep{miluzio:2010tl}.  The object was not detected on 2009 Aug. 25 (limit $K=19.0$ mag).
Based on a spectrum taken on 2010 Sept 8, the SN was classified as a type II event more than 30 days past maximum light  \citep{marion:2010sz}. We obtained a follow-up spectrum on 2010 Sept 15 with EFOSC2 at the NTT showing the typical features of type II SN, that is H  Balmer lines,  NaI D doublet, the NIR CaII emission triplet along with a number of FeII lines in the blue region.
The GELATO spectral comparison tool found a best  match with the type IIP SN~1999em \citep{elmhamdi:2003vn} at about +60 days, adopting a reddening of about $A_B=0.5$ mag (Fig.\ref{spec}). This is fully consistent with the color curves comparison between the two SNe.

\item[{\bf SN 2011ee}] was discovered on 2011 June 27.3 UT  \citep{miluzio:2011kg}. 
The object was not detected on a K-band image taken on 2010 September 7 ($ K>19.0$ mag). 
An optical spectrum was obtained  with X-Shooter at the VLT  on 2011 July 17.3 UT showing that the 
transient is a type Ic SN.  The GELATO code finds   a best match with SN~2007gr \citep{hunter:2009hv} at maximum (Fig.~\ref{spec}). Because  the classification of type Ic SNe near maximum is sometimes ambiguous we
obtained a followup spectrum 
about two months later (2011 Sept 20) using OSIRIS at  the GTC. Again the spectrum
is very similar to SN~2007gr at corresponding phase (Fig. \ref{spec}  bottom panel).
The color comparison and the 
spectral match are consistent with a negligible host galaxy extinction.

\item[{\bf PSN2010 in IC~4687}] was discovered on 2010 May 21.3 UT  in the northern component of a galaxy triplet that include also IC~4686 and, 1 arcmin to the south of IC 4687, IC~4689. IC~4687 has a chaotic structure formed byf stars, gas and dust and a large curly tail.  The transient was not detected on a K-band image taken on 2009 Aug. 8.1 ( $K>19.0$ mag). We obtained an optical/infrared spectrum   with X-Shooter at VLT  on 2010 June 5. However, because of its very low S/N, we could not derive a convincing classification and  therefore we had to rely on the  K-band photometry. Comparing the K band absolute light curve of PSN2010 ($A_B ({\rm host}) = 0$) with template light curves of different SN types  we found a good match with SN  2005cs a prototype of under-luminous type IIP SN \citep{pastorello:2009sy},  assuming that the detection of PSN2010 was 2 months after the explosion. However, lacking color measurements, we could not constraint the extinction and indeed, assuming a high extinction  $A_B \sim 8$ mag, we found an alternative good match with the light curve of SN~1999em (Fig.~\ref{psn2010_lc}). Intermediate values may also be adopted by fitting other SN II.

\begin{figure}
\centering
\includegraphics[width=.45\textwidth]{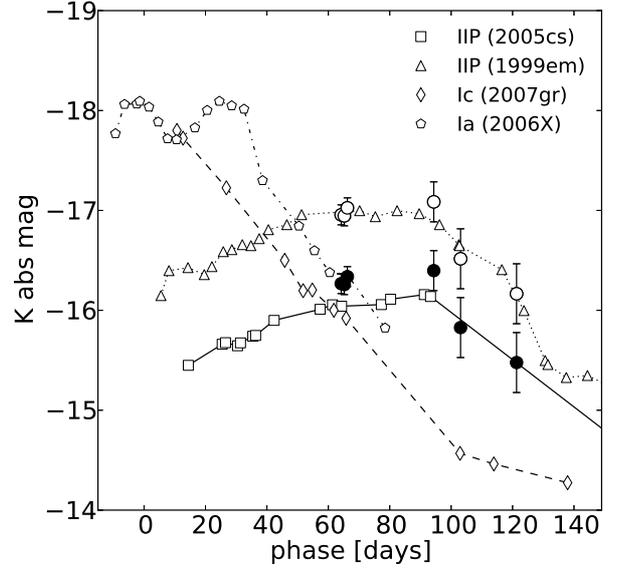}
\caption{K band absolute light curve of the PSN2010 (black dots), compared with those of template SNe. The empty circles show the same data but assuming an extinction $A_B = 8$ mag. } \label{psn2010_lc}
\end{figure}

\item[{\bf PSN2011 in IC 1623}] was discovered on 2011 July 21.4 UT in the western component of a galaxy pair.   The object was not  detected on a K-band image taken on 2010 Sept. 5 ($K>19.0$ mag). Unfortunately, due to bad weather in the scheduled nights, we could not obtain a spectroscopic observation of the transients. We have to rely on  three epochs of photometry,  in K complemented by two epochs in the optical R and I bands. A simultaneous comparison of the absolute observed luminosity with template SNe give a best fit with the SN~Ic 2007gr one month after maximum
(Fig.~\ref{psn2011_lc}). Assuming this classification, from the colors we can constrain the extinction to be $A_B = 0.5 \pm 0.5$ mag. However, we have to admit that, within the errors, the photometry of PSN2011 can be consistent also with a type IIP at about 3 months after explosion.

\begin{figure}
\centering
\includegraphics[width=.45\textwidth]{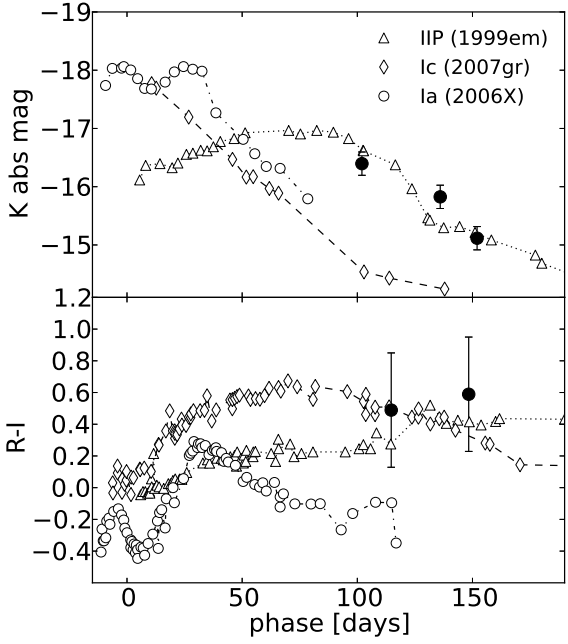}
\caption{K and  R-I colours  of  PSN 2011, compared with those of template SNe.
For this plot we adopted $A_B = 0.5$ mag, which improves the fit of the R band photometry ($\Delta R = 0.3$ mag).
} \label{psn2011_lc}
\end{figure}
\end{description}

To recap, during the search we discovered 6 SNe. Four received a spectroscopic classifications: one as a type Ia and three as core collapse events, a type IIn, a type IIP and a type Ic. For the other two, based on the sparse photometry, we argue that most likely they are 
 core collapse SNe, with a best fit as  type IIP and type Ic, respectively.

\subsection{Search detection limit}\label{artstar}

In order to derive the SN rate from the number of detected events it is crucial to obtain an accurate estimate of the magnitude detection limit for each of the search images and for different locations in the images. As it has been shown in Fig.~\ref{dseeing}, the detection efficiency is influenced by the sky conditions at the time of observations (namely seeing and transparency) and by the transient position inside the host galaxy.

The magnitude limit for SN  detection has been estimated through artificial star experiments. 
The procedure we adopted was the following:

\begin{enumerate}
\item fake SNe of different magnitudes are simulated with the PSF derived from isolated field stars;
\item the image is segmented in a number of intensity contour levels. We took  denser contours in the nuclear regions because the magnitude limit changes rapidly with background intensity;
\item one fake SN of specific magnitude is randomly placed  inside a chosen intensity contour;
\item the image with the fake SN is processed through the image difference and transient detection pipeline;
\item if the fake SN results in a detected transient, the experiment is repeated with a fainter artificial star until we have a null detection. The fainter magnitude for which the fake SN is detected defines the magnitude limit for the given background intensity level;
\item Steps 2 to 5 were repeated for each contour level three times to enhance the statistical significance of the results. The average value for each contour level has been adopted as the magnitude discovery limit for the given background intensity.
 \end{enumerate}

\begin{figure}
\centering
\includegraphics[width=.45\textwidth]{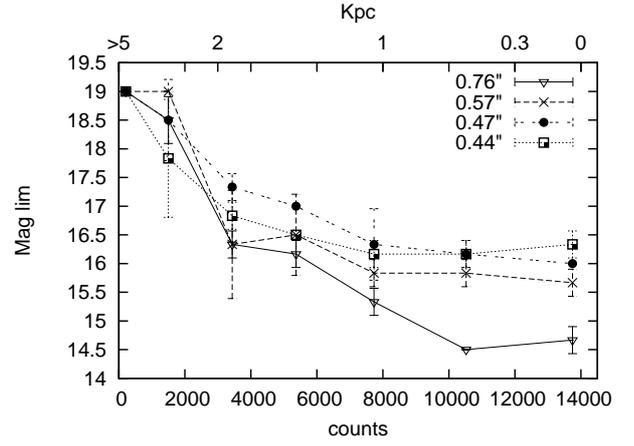}
\caption{Magnitude limit vs. background intensity for four different observations of NGC~7130. The points are labeled with the respective image seeing. Errorbars show the range of the magnitude limits  from the three different experiments. } \label{art1}
\end{figure}

To illustrate the results, a plot of the magnitude limit versus background counts for four observations of the galaxy  NGC~7130  is shown in Fig.~\ref{art1}. Each epoch is labelled with the image seeing, while the errorbar shows the range of limiting  magnitudes for the three experiments. The top x-axis shows the linear distance in Kpc from the galaxy center.   

It can be seen that, as expected,  the magnitude limit is lower in the nuclear regions  which, for a typical galaxy, correspond to 1.5-2.0 kpc. Epochs with different seeing have similar magnitude limits in the galaxy outskirts (typically $K~\sim 19$ mag), while in the nuclear region when seeing is poorer the magnitude limit is brighter (in the worst case even 5-6 mag brighter than in the galaxy outskirts). 
 
\section{SN search simulation}\label{montecarlo}

To evaluate the significance of the detected events we elaborated a simulation tool that returns the number and properties of   expected events based on   specific features of our SN search, a  number of parameters describing our current knowledge of SBs and SN properties. The tool uses a MonteCarlo approach which simulates the stochastic nature of SN explosions.   By collecting  a number of MonteCarlo experiments with the same input parameters, we can test whether the observed events are within the expected distribution. On the other hand by varying some of the input parameters,  we can test the influence of specific assumptions.

\subsection{The simulation tool}

Our MonteCarlo (MC) simulation tool is  built in a {\em Python} environment and makes use, for the different inputs, of standard values taken from the literature. For those that are more controversial, we will give references with some discussion.
 The basic ingredients of the simulation are:

\begin{itemize} 
\item relevant data for the selected SBs, namely:  redshift, galactic extinction, infrared fluxes  from IRAS catalogues  at 25, 60 and $100\mu$,  B magnitude corrected for internal extinction\footnote{The galaxy internal extinction was retrieved from HyperLeda \citep{paturel:2003fk}}, Hubble morphological type. These data were retrieved from NED;
 \item information describing the SN properties  for each of the SN types considered here: SNe Ia and core collapse events, including SNe
  IIP, IIL, IIn and Ib/c. K band  template  light curves were constructed starting from B template  light curves \citep{cappellaro:1997mz} and $B-K$ k-corrections for the given galaxy redshift \citep{botticella:2008fr}.  We note that the results of this procedure are in close agreement  with the K template light curves of \cite{mattila:2001gf}. For the SN luminosity functions we adopted as reference those of \cite{li:2011jf}, but we also tested for the possible presence of a significant population of faint core collapse which may be suggested by the analysis of very nearby SNe \citep{horiuchi:2011xv}.
From  the LOSS project we adopted also the relative rates for the different SN types \citep{li:2011qf};
\item details of the search campaign: log of observations, magnitude detection limit for each observation as a function of the host galaxy background intensity;
\item the number of SNe expected from a given star formation episode. For core collapse SNe, this is determined only by the adopted mass range of the progenitors and IMF slope. In fact, for our purposes, we can neglect the very short time delay from CC progenitor  formation to explosion. For type Ia  SNe we need to consider the realization factor, that is the fraction of events in the proper mass range which occurs in suitable close binary systems and the delay time distribution (Sect.~\ref{snsfr}).;
\item the  depth and distribution of the extinction by dust inside the parent galaxies (cf. Sect.~\ref{extinction});
\item the  star formation spatial distribution in the parent galaxies (cf. Sect.~\ref{sfrdistribution}).
\end{itemize}

Hereafter, we discuss our assumptions about the parameters of the simulation.

 \subsubsection{From IRAS measurements to Star Formation Rate}\label{TIRdef}

The SFR in SBs can be estimated on the basis of the galaxy  total infrared luminosity (L$_{TIR}$) under the assumption that dust re-radiates a major fraction of the UV luminosity, and after calibration with stellar synthesis models. In turn the TIR luminosity can be estimated from FIR flux measurements.

 \cite{helou:1988br} provided a prescription for deriving the FIR emission from IRAS measurements:

$$FIR = 1.26 \times 10^{-14}\, [2.58 \, f_\nu(60\mu m) + f_\nu(100 \mu m)]$$

where FIR is in W m$^{-2}$  and $f_{\nu}$ are in Jansky. FIR fluxes are converted into TIR fluxes by using  the  relation of  \cite{dale:2001vy}

$$ \log \frac{TIR}{FIR} = a_0+a_1 x + a_2 x^2 + a_3 x^3 + a_4 x^4$$

where $x = \log \frac{f_\nu(60\mu m)}{f_\nu(100 \mu m)}$ and $[a(z=0)]\simeq[0.2378, -0.0282,\\ 0.7281, 0.6208, 0.9118]$\\

TIR fluxes are converted in luminosities using the adopted distances:

$$ L_{TIR} = 4\pi D^2 \,TIR $$

Finally, the relation between the SFR ($\psi$), and  L$_{TIR}$ was derived by \cite{kennicutt:1998tf} from  SB galaxy spectral synthesis model adopting 10-100 Myr continuous bursts and a Salpeter IMF as: 

$$ \frac{\psi}{[{\rm M}_\odot\, {\rm yr}^{-1}]} =  \frac{L_{TIR}}{2.2 \times 10^{43} ~~ [{\rm erg}\,{\rm s}^{-1}]} = \frac{L_{TIR}}{5.8 \times 10^9 ~~  [{\rm L}_\odot]}$$

\subsubsection{SNR and SFR}\label{snsfr}

In general,  the rate of SNe  expected at a specific time, $\dot{n}_{SN}(t)$, for a stellar population depends on the star formation history,  the number of SNe per unit mass from one stellar generation  (labelled as SN  productivity)  and the distribution of delay time from star formation to explosion for the specific SN type. Following the notation of  \cite{greggio:2005ph,greggio:2010pd}: 

\begin{equation}\label{rate}
\dot{n}_{SN}(t)=\int^t_0\psi(t-\tau) \, k_{SN} \, f_{SN}(\tau) \, d \tau
\end{equation}

where $\psi(t)$ is the star formation rate, $f_{SN}$ is the distribution of the delay times $\tau$ and $k_{SN}$ is the supernova productivity. The equation shows that at a fixed epoch $t$ since the beginning of star  formation, the rate of SNe is obtained by adding the contribution of all past stellar generations, each of them weighted with the SFR at the appropriate time.

\subsubsection*{Core Collapse SNe}

For core collapse SNe the delay time from star formation to explosion (2.5 Myr for 120 M$_\odot$ stars up to 40 Myr for  8 M$_\odot$ stars) is  short compared with the typical SBs duration \citep[$200-400$ Myr,][]{mcquinn:2009uq}.  Assuming that the SFR in the SB was constant during the past 40 Myr, the expected CC SN rate, $\dot{n}_{CC}$, is proportional to the current SFR:

$$  \dot{n}_{CC} = k_{CC} \times \psi $$

The supernova productivity $k_{CC}$ is derived by integrating the IMF, $\phi(m)$, and assuming a CC progenitor mass ($M_{CC}$) range:

$$ k_{CC} = \frac{ \int^{M_{CC}^U}_{M_{CC}^L} \phi(m) dm}{\int ^{M_{U}}_{M_{L}} m \phi(m) dm} $$

where $M_{CC}^L$ and $M_{CC}^U$ are respectively the lower and upper mass limits for SN~CC progenitors and $M_L$, $M_U$ are the lower and upper stellar mass limit. To be consistent with the Kennicutt's SFR calibration we adopted a Salpeter IMF, that is: 

$$\phi(m) \propto m^{-\alpha}     ~~~~~ \mbox{with} ~~ \alpha=2.35  ~\mbox{and}~ 0.1\,{\rm M}_\odot<M<100\, {\rm M}_\odot $$

Assuming $8 < M_{CC} < 50 \,{\rm M}_{\sun}$ for the CC progenitor mass range, $k_{CC} = 0.007\,{\rm M}^{-1}$. This number changes significantly if we adopt a different IMF, e.g. $k_{CC} = 0.011$ for a Kroupa IMF or $k_{CC} = 0.039$ for an extreme {\em Starburst} IMF \citep{dwek:2011lq}. We soon note however that, because the IMF enters also in the conversion from $L_{TIR}$ to $\psi$, the expected rate of SN events is almost independent on the selected IMF (cf. Sect.~\ref{tests}) provided the choice is consistent.

More important is the assumption on the mass range for  CC progenitors which is not well constrained. Actually, while changing the upper limit of the progenitor mass from $40$ to $100\, {\rm M}_{\sun}$ makes a modest  10\% increase in the CC SN productivity,  the lower mass limit is crucial, with $k_{CC}$ decreasing by 30\% if we adopted  $M_{CC}^L=10\,{\rm M}_{\sun}$ instead of the favored value of  $8\,{\rm M}_{\sun}$ \citep{smartt:2009mq}.

\subsubsection*{SN~Ia}

Estimating the expected rate of SN~Ia is complicated because the delay time distribution $f_{Ia}$, while still uncertain, certainly ranges from short to very long time. In particular it has been suggested that SN~Ia can be divided into two classes, one with a short delay time whose rate scales with the current SFR (also called {\em prompt}), and a second with a long delay time ({\em tardy}),  whose rate scales with the average of the SFR along the entire galactic evolution \citep{scannapieco:2005rr,mannucci:2006zi}. 
While stellar evolution arguments  \citep{greggio:2010pd, greggio:2005ph, greggio:1983gr} and more recent data  \citep{maoz:2012uq, totani:2008tm} suggest a continuous distribution of the delay time instead of two distinct classes, the schematization is still a fair approximation that help in simplifying the problem of predicting the expected SN~Ia rate in SBs.

In  general, for a galaxy of the local Universe, $\sim13$ Gyr after the beginning of SFR, we can identify the contribution of the two components as follows \citep{greggio:2010pd}:

\begin{equation}\label{nia}
\dot{n}_{Ia}(13)=k_{Ia} \times \left( \psi_C \int_0^{0.1} f_{Ia}(\tau)\,  d\tau \, +\,  \psi_P \int_{0.1}^{13}  f_{Ia}(\tau)\,  d\tau 
\right)
\end{equation}

where $\psi_C$ and $\psi_P$ are the average SFR over, respectively, the last 0.1 Gyr (current SFR) and from 0.1 to 13 Gyr ago (past SFR).  The SN productivity $k_{Ia}$ is the product of  the number of stars per unit mass in the adopted progenitor mass range (0.021 for a Salpeter IMF and a mass range $3{\rm M}_{\sun}<M<8\, {\rm M}_{\sun}$) and the realization fraction, the actual fraction of systems which make a successful explosion ($\sim 5\%$ according to the most recent estimate) \citep{maoz:2012kx}. 
We assume that SF history in SBs can be described schematically with two components: a constant SFR during the  galaxy evolution which created  the galaxy stellar mass, and an on-going episode of intense SFR which is the source of the strong TIR emission. Neglecting the contribution of the ongoing SB to the galaxy stellar mass, we can approximate $\psi_P \simeq M/13\times10^9$, and  write Eq.~\ref{nia} as follows:

$$\dot{n}_{Ia}(13) \simeq k_{Ia} \left(  \psi_C F_{Ia}^p \, +\,  \frac{M}{13\times 10^9} F_{Ia}^t \right)$$

 where $F_{Ia}^p =  <f_{Ia}^p> \times 0.1$ and $F_{Ia}^t = <f_{Ia}^t> \times 13 \simeq 1- F_{Ia}^p$ are the relative fraction of {\em prompt} and {\em tardy} events derived by integrating the delay time distribution in the relevant time range.  In our approximation $\psi_C$ can be derived from the observed $L_{TIR}$ and the galaxy mass from the $K$ magnitude and $B-K$ colors \citep[cf.][]{mannucci:2005mb}. 

The relative contribution of the two SN~Ia components has been a very debated issue in the last few years, ranging from $F_{Ia}^p \sim 50\%$ \citep{mannucci:2006zi} to $F_{Ia}^p \sim 10\%$  from standard stellar evolution  scenarios \citep{greggio:2010pd}. In our simulation we adopted as reference an intermediate value, $F_{Ia}^p \sim 30\%$.

\subsubsection{Extinction}\label{extinction}
 
Dust extinction in SBs is very high, especially in the nuclear regions. For instance, \cite{shioya:2001cy} 
found that fitting the spectral energy distribution of the nuclear region of Arp~220 requires a visual extinction $A_V> 30$ mag. Actually,  according to \cite{engel:2011ys}, "over most of the disk the near-infrared obscuration is moderate, but increases dramatically in the central tens of parsecs of each nucleus". Similar high extinction, $A_V \sim 20$, was found for the SB region of Zw~096 \citep{inami:2010vn}. 

As a first order approximation, for our simulation we assumed  that the  extinction  has the same distribution of the SF (see next section) with a maximum value $A_V = 30$ mag corresponding to the SFR peak and scaled linearly in the other regions. While this is a crude approximation, it turns out that the actual choice of extinction correction has little impact for our simulation. In the nuclear, high extinction regions the SN detection is limited by the reduced performance of the image subtraction  algorithm in these high surface brightness regions. At the same time, our IR search is largely insensitive to variation in the (moderate) extinction of the outer galaxy regions. 
 
For the wavelength dependence of extinction we adopted the Calzetti's law with $R_V = 4.05 \pm 0.8$  \citep{calzetti:2000ht}.

\subsubsection{Star Formation Distribution}\label{sfrdistribution}

The spatial distribution of the SFR is a key ingredient of the simulation. This is because we expect that SNe occur more frequently in the high SF regions where, on the other hand, our detection efficiency is lower. In principle, the FIR emission which is used to estimate the SFR would also be  a good tracer of its spatial distribution. However, it turned out that the available  MIR imaging for the galaxies of our sample (mainly obtained with the Spitzer observatory) do not have enough spatial resolution for  mapping  the compact SB structures. 

Selected K-band  images from our survey can have excellent resolution but,  as  is well-known, the near IR emission better traces the old star population, that is the galaxy mass distribution more than the SFR distribution. Therefore for an estimate of the SFR  concentration, we are forced to an indirect, statistical approach.
 
Our starting point is the SB classification by \cite{hattori:2004ye}, who derived a correlation between the global SBs properties, such as FIR colors,  and the compactness of the SF regions. These range from  very compact  ($\leq$100 pc) nuclear starbursts with almost no star-forming activity in the outer regions (type 1), to extended starbursts with relatively faint nuclei (type 4), with type 2 and 3 as intermediate cases. In addition, they found a trend for  galaxies with more compact  SF region showing a higher star formation efficiency and  hotter far-infrared color.  They also found that the compactness of SF regions  is  weakly correlated with the galaxy morphology, with disturbed objects showing preferentially more concentrated SF.  On the other hand, an appreciable fraction ($\sim50\%$) of their galaxy sample was  dominated by extended starbursts (type 4).
The significant variations in the degree of concentration of  the SB SF regions has been recently confirmed by  \cite{mcquinn:2012zr}.

In an attempt to characterize the SF spatial distribution for the SBs of our sample we derived estimates of their morphological class and FIR colors.  In particular, following \cite{hattori:2004ye}, SBs with strong tidal features and a single nucleus were classified as ''mergers'' (M), galaxy pairs with an overlapping disk or a connecting bridge were classified as ''close pairs'' (CP) if the projected separation is $< 20$ Kpc and galaxies that have a nearby ($<100$ Kpc) companion at the same redshift were classified as ''pairs'' (P). The remaining objects were classified as ''single'' (S). The classification of the SBs of our sample is listed in Tab.~\ref{classification} along with the galaxy FIR colors, $\log f_{60} / f_{100}$, $\log f_{25}/f_{60}$.
 
\begin{table}
\begin{center}
\caption{Morphological classification, FIR colors and compactness classification for the SBs of our  sample. Following \cite{hattori:2004ye}, type 1 have SF region $< 500$ pc, type 2 $<1$ Kpc, type 3 $> 1$ Kpc and type 4 have extranuclear star formation.}\label{classification}
\begin{tabular}{lccccc}
\hline\hline
Galaxy          & Morp.      &$\log f_{60}/f_{100}$ & $\log f_{25}/f_{60}$ & Comp. \\
                &    class           &           &                   &          class       \\
\hline 
CGCG011-076       & S                & -0.20         &  -0.89  &  3\\
CGCG043-099       & S                & -0.19         &  -1.05  &  3 \\ 
ESO148-IG002      &CP               &  0.01         &  -0.82  &  2\\
ESO239-IG002      &M                 & -0.04         &  -0.79  &  2 \\ 
ESO244-G012       &CP               & -0.10         &  -0.68  & 2 \\ 
ESO264-G036       &S                  & -0.34         &  -0.96  & 4 \\ 
ESO286-IG019      &M                 &  0.07         &  -0.81  & 2 \\
ESO440-IG058      &P                  & -0.23         & -0.99  & 3 \\ 
ESO507-G070       &S                  & -0.08         & -1.21  & 1 \\ 
IC1623A/B         &CP               & -0.14         & -0.80  & 2\\
IC2545            &M                 &  0.01          & -0.88  & 2 \\ 
IC2810            &P                  & -0.22          & -1.00  & 3\\ 
IC4687/6         &P                  & -0.13           & -0.76 & 2\\
IRAS12224-0624  &S                 & -0.14           & -1.48 & 1 \\
IRAS14378-3651  & S                & -0.08           & -1.00  & 1 \\
IRAS16399-0937  &CP              & -0.24           & -0.87  & 3\\ 
IRAS17207-0014  &M                 & -0.05           & -1.36 & 1\\ 
IRAS18090+0130  &P                 & -0.31           & -0.98 & 4\\
MCG-02-01-051/2 &P                  & -0.14           & -0.79 & 2\\ 
MCG-03-04-014   &S                   & -0.15            & -0.91 & 3\\ 
NGC0034         &M                        & 0.01             & -0.85 & 2\\ 
NGC0232         &P                        & -0.23             & -0.89 & 3\\
NGC3110         &P                       & -0.30             & -1.00  & 4   \\ 
NGC5010         &S                       &-0.33              & -0.85  & 4   \\ 
NGC5331         &CP                   &-0.29               &-1.0 0  &  4  \\   
NGC6240         &CP                   &-0.06               & -0.81 &  2  \\
NGC6926         &S                      &-0.31              & -0.83  &  4   \\ 
NGC7130         &P                     &-0.19                & -0.89&3\\
NGC7592         &CP                  &-0.12                &-0.92&2\\
NGC7674         &P                   &-0.19                &-0.45 & 2 \\ 
\hline
\end{tabular}
\end{center}
\end{table}

\begin{figure}
\centering
\includegraphics[width=.45\textwidth]{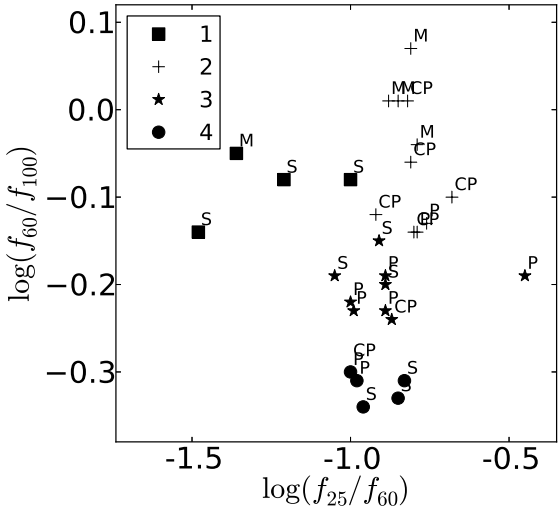}
\caption{FIR color of the SBs of our sample. The different symbols identify the compactness class \citep{hattori:2004ye} while the label show the morphological type.  } \label{fircol}
\end{figure}

We attributed to each galaxy a compactness class  on the basis of its correlation with the FIR colors as shown in Fig.~4 of \cite{hattori:2004ye} that for the object of our sample corresponds to our Fig.~\ref{fircol}.  As it can be seen, we also confirmed their claim of a (weak) relation of FIR color and, as a consequence, compactness class with SB morphology.

The next step is based on  \cite{soifer:2000zn,soifer:2001ly}. For a number of SBs galaxies they  plotted the  MIR and NIR emission curve of growth finding that in general the MIR emission is more concentrated, while  only for few galaxies the MIR and NIR curves of growth show  a similar trend.
Actually we found that, to a first order approximation, the MIR emission profile of a given galaxy can be matched by  NIR profile powered to an exponent $\alpha$ which ranges between 1, when the two profiles are similar, to 2, when the MIR emission is strongly concentrated. When we classify  the same galaxies  with the compactness criteria of \cite{hattori:2004ye}, we found that (as expected)  the galaxies  with compact SF regions (type $1-2$)  are characterized by more concentrated MIR  emission ($\alpha=2.0-1.5$, respectively), while galaxies with  extended SF region (type 3-4) have  similar MIR vs. NIR profiles ($\alpha=1.25-1.0$, respectively).
As a reference, we notice that in the typical case of NGC 6240, assuming $\alpha=1$ corresponds to locate  50\% of the SFR within 1.5 Kpc, whereas for $\alpha=2$  the same SFR fraction is enclosed within  500 pc.

As a result of this discussion we have a prescription to estimate the SF distribution based on the observed $L_K$  map, and adopting a power index $\alpha$ appropriate for the compactness class of the given SB galaxy (Tab. \ref{refvalue}).

\subsubsection{Flow chart of the simulation}

Having defined all the ingredients of the simulation, we can now describe how this proceeds. The simulation flowchart can be summarized as follows:

\begin{enumerate}
\item for each  galaxy of the sample, based on the estimated total SFR and adopted progenitor scenarios, we  compute the expected number of SNe per year;

\item a time interval is chosen so that 100 SNe are expected to explode in the given galaxy in that period.  
The time interval ends  with the last observations of the galaxy. Given the expected SN rates it is for all galaxies  much longer than the duration of our monitoring campaign. The reason to simulate 100 events is to avoid having to deal with fractional SN numbers for the different subtypes.
We assign to each event a random epoch of explosion chosen within the defined time interval;

\item  each SN is assigned to a random explosion site  inside the parent galaxy according to the SFR spatial distribution;

\item a peak magnitude is also assigned  to each SN,  with a random value  derived from  the adopted SN luminosity function for the specific subtype.  We also associate  to the event an extinction value  randomly extracted  from a  gaussian distribution whose mean value  depends on the position of the SN,  and $\sigma=1/3$ of the mean value;

\item the apparent magnitude is thereafter calculated at the epochs of available observations, knowing the galaxy distance modulus and SN epoch of explosion. This is compared with the search magnitude detection limit for the given position and, when brighter, the SN is added to the simulated discovery list.
\end{enumerate}

The process, iterated for all the galaxies of the sample, defines a single simulation run. 
Outcomes of the simulation are the expected number of SN discoveries, their types, magnitudes, extinctions and positions inside the host galaxies. To explore the distribution of the outcomes from the random process,  a complete experiment is made by  collecting a minimum of a  hundred single simulation runs.  

For the reference simulation we used theinput parameters summarized in Tab.~\ref{refvalue}. In the next section we will compare the prediction of the simulation with the current SN
discoveries.

\begin{table}
\begin{center}
\caption{Input parameters for the reference simulation}\label{refvalue}
\begin{tabular}{cc}
\hline 
$L_{TIR}/SFR$ calibration & \cite{kennicutt:1998tf} \\
\\
IMF                                          &  Salpeter                        \\
\\
SFR distribution                    & $\propto L_K^{\alpha}$\\
                                                 & for compactness class 1,2,3,4 \\
                                                & $\alpha=2.0,1.7,1.25,1.0$\\
                                                \\
A$_V$                                    & $\propto SFR$ ~(max 30 mag)\\
\\
CC mass range                  & $8-50 {\rm M}_{\odot}$\\
Ia   components                  & 30\% prompty, 70\% tardy\\ 
\\
SN luminosity function        & LOSS \citep{li:2011jf}\\
\hline
\end{tabular}
\end{center}
\end{table}

\section{Comparison between observed and expected SN discoveries}\label{results}

\begin{figure}
\centering
\includegraphics[width=.45\textwidth]{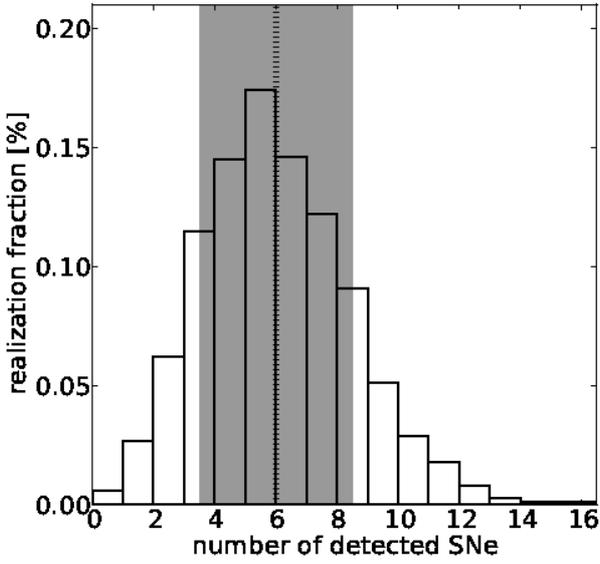}
\caption{Histogram of the number of expected SNe from our MonteCarlo experiments. The dashed vertical line indicates the number of observed events and the grey area its 1$\sigma$-Poissonian uncertainty.} 
\label{histnumb}
\end{figure}

As we  outlined above, from a large number of MonteCarlo simulation runs we obtain the distribution of the  expected SN discoveries.  This is shown in  Fig.~\ref{histnumb}, where each  bin of the histogram is the predicted probability of observing the specific number of SN discoveries whereas the dashed line marks the number of actual SNe discovered and the shaded area  shows its  1-$\sigma$ Poissonian uncertainty range.

We found that with the adopted simulation scenario and input parameters we should have expected, on average, the discovery of $5.3\pm2.3$ SNe. In 68\% of the experiments (1-$\sigma$) the expected number is in the range 4-8 which is in excellent agreement with the observed number of 6 events. 

The prediction of the simulation is that  almost all SNe are CC (5.1 SN~CC vs. 0.2 SN~Ia), though in $10\%$ of the experiments at least one type Ia is found (that is what we have from the real SN search).

\begin{figure}
\centering
\subfloat[Expected (line-only) and observed (grey) K magnitude distribution at the discovery]{\includegraphics[width=.35\textwidth]{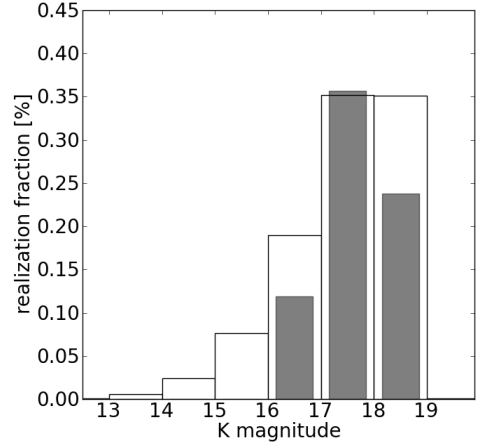}}\\
\subfloat[Expected (line-only) and observed (grey) extinction distribution. In light grey we indicate  the allowed range for the extinction of PSN2010 (see text).]
{\includegraphics[width=.35\textwidth]{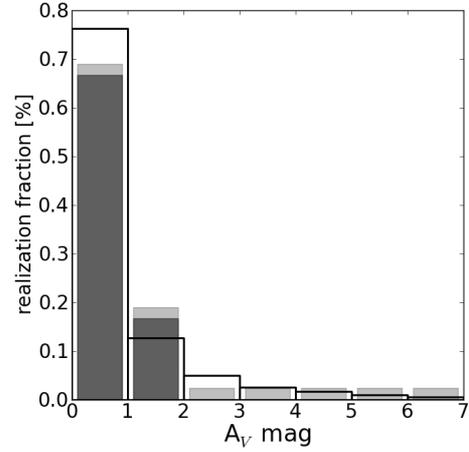}}\\
\subfloat[Radial distribution  of expected (line-only) and detected (dark grey) SNe.
For regular galaxies the surface brightness decrease monotonically with  radial distance (the upper axis  shows this correspondence for one of the galaxy of our sample). In light grey we show the distribution of injected artificial SNe (see text).]
{\includegraphics[width=.35\textwidth]{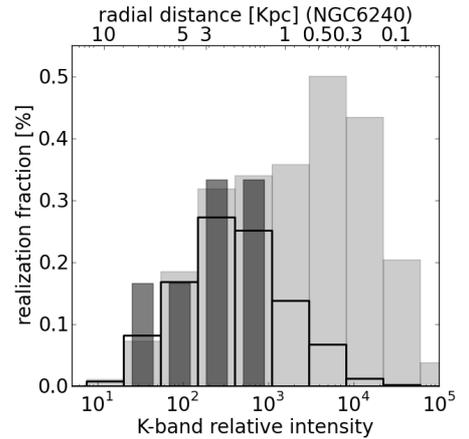}}
\caption{SN properties comparison}\label{snproperties}
\end{figure}

The distributions of some of the expected and observed SN properties are compared in Fig.~\ref{snproperties}. For the simulation, we show the distribution across a large number of experiments (line-only histogram) while the grey shaded histogram represents the actual observations. 

The top panel in Fig.~\ref{snproperties} shows the distribution of  the apparent magnitudes at the discovery. The good agreement between simulations and observations is  a crucial consistency check of our estimates of the magnitude detection limit: if the discovered SNe were systematically fainter/brighter then expected, this would indicate, respectively,  an underestimate/overestimate of the search detection efficiency. 

A comparison of the simulated vs observed extinction distribution is shown in the middle panel of Fig.~\ref{snproperties}. For  the observed distribution the case of PSN2010 for which extinction is ambiguous is shown in light grey. Again the simulation is in good agreement with the observations. This argues in favor of the consistency of the input assumptions. The fact that, in our IR search we expect that most of  detected  SNe have low extinction ($\sim 75\%$ with $A_V < 1$ mag) is a consequence of  our assumption that the extinction is very high in the nucleus and  rapidly decreases with the galaxy radius, following the same trend of the SFR. This does not means that extinguished SN  are intrinsically rare but that they are confined to the galaxy nuclear regions where extinction is extremely high even in the IR (see next). At the same time we can exclude the presence of a significant population of SNe with intermediate extinctions: we would easily detect them in our infrared search.\\
 \cite{mattila:2001gf} found an average value of $A_V$=30 mag for the extinction towards the SN remnants of M82. Confirming the presence of high extinction (about $A_V\sim$15-45 mag) in the innermost 300 pc regions. On the other hand, \cite{kankare:2008ys} found an host galaxy  extinction of $A_V\sim$16 mag for the SN 2008cs, located at about 1.5 Kpc, relatively far from the galaxy nucleus.

Finally, in the bottom panel  of Fig.~\ref{snproperties} we compare the distribution of locations inside the host galaxy for the expected (line-only) vs observed (dark grey) SNe. The different location are identified by the K band pixel counts: in general high counts occurs in the nuclear regions while low counts are in the outskirts (we use pixel counts instead of radial distances because the latter is difficult to be defined for  galaxies with irregular morphology or double nuclei. However a indicative correspondence from pixel count to radial distance is shown it the top axis of the figure for a galaxy with regular morphology. There is a mild indication of a deficiency of observed events in  regions with high pixel counts. Taken to face value this may suggest  a minor overestimate of the detection magnitude limit in the nuclear regions. Given the poor statistics we cannot derive definite conclusions and therefore we will not elaborate further this issue.

In the same figure we show also (in light grey) the distribution of locations  of the expected events for an ideal case where the magnitude detection limit in the nuclear regions is as deep as in the outskirts, and extinction is negligible. The experiment shows that the fraction of events that remains hidden to our search in the galaxy nuclear regions due to the combined effect of reduced search efficiency and high extinction is very high, being about $60\%$ \citep[cf.][]{mattila:2012er}.

\subsection{Uncertainties and effect of different assumptions}\label{tests}

\subsubsection*{Magnitude detection limit}

One of the main source of uncertainty for the simulation is related to the estimate of the magnitude detection limit, $mag_{lim}$. For the reference simulation, we adopted as $mag_{lim}$ the mean value out  of three artificial star experiments conducted for a number of selected positions inside the host galaxy (cf.~\ref{artstar}). The dispersion of measurements, that is the uncertainty on $mag_{lim}$, is quite large with a typical range of $\sim 0.5$ mag but, in more extreme difficult cases,  it can be as large as 2 mag.

To test the propagation of this uncertainty,  we performed MonteCarlo experiments assuming alternatively the lower and higher $mag_{lim}$ out of the three experiments. We found that the predicted number of SNe is respectively $6.2\pm2.5$ and $4.7\pm2.2$, that are $+17\%$ and $-11\%$  with respect to the numbers from the reference simulation. The fact that the error is significant is the  reason why we spent a significant effort for a detailed estimate of the detection limit.

\subsubsection*{SN Luminosity Function} 

In the reference simulation we use a  gaussian distribution for the SN luminosity function (SN-LF)  with a mean value and dispersion taken from \cite{li:2011jf}. However, \cite{horiuchi:2011xv}, based on a small sample of very nearby  SNe, claimed that the faint end of the SN-LF is underestimated and SN~CC fainter than ${\rm mag}\simeq -16$ could made up to 50\% of  the distribution, to be  compared with  20\%  of the sample of \cite{li:2011jf}. 
 On the other hand, \cite{mattila:2012er} argued that \cite{horiuchi:2011xv}  overestimated the fraction of intrisically faint CCSNe since they neglect the host galaxy extinction for their SN absolute magnitudes. We performed a MonteCarlo experiment adopting the Horiuchi's SN-LF  and found that in this case the expected number of events would be low, only 3.3 on average.
This is because most faint events are expected to fall below the search detection limit. The fact that the actual discoveries are twice this number argues against a large fraction of faint SN-CC
\citep[cf.][]{botticella:2012sh} 

\subsubsection*{IMF and SN CC progenitor mass range} 

The IMF enters both in the estimate of the number of SN progenitors and in the calibration of TIR luminosity in terms of SFR relation.  However, the expected rate of CC SNe in our
sample is  virtually independent of the IMF slope. Indeed, for a given total mass of the
parent stellar population, top heavy IMFs imply both a higher number of CC progenitors as well
as a larger luminosity. Following \cite{dwek:2011lq} the number of CC progenitors per unit mass is $k_{CC}=  0.007, 0.011$  and 0.039 ${M_\odot}^{-1}$ respectively for a Salpeter, a Kroupa and a Starburst IMF , assuming that the progenitors range from 8 to 50 $M_{\sun}$. At the same time
the total luminosity of a SB forming stars with a SFR of 1 $M_{\sun}\,{\rm yr}^{-1}$ over a period of
10 Myr (i.e. a  $10^7$ $M_{\sun}$ stellar population) is  $4.71 \times 10^9$, $7.33 \times 10^9$ and $2.55 \times 10^{10}$ L$_{\odot}$ again for a Salpeter, a Kroupa and  a Starburst IMF, respectively. The M/L ratio of such SB is then 0.0021, 0.0014 and 0.0004 (solar  units) for the three IMFs, and the expected number CC SNe originating from it is $\simeq 1.5$ every $10^{5} L_\odot^{-1}$  for all the three IMFs.
Working out the numbers, it turns out that the  SN~CC rates from a population with a given $L_{TIR}$ is almost independent on  the IMF, provided a consistent choice is made. 

Crucial is instead the assumption of the SN~CC  progenitor mass range, in particular the lower limit.
Indeed if we adopted an upper limit of $100\, {\rm M}_{\sun}$ instead of the reference value of 50 M$_{\odot}$ the expected number of SNe  would be $5.5\pm2.1$,  only $\sim$5\% higher then  the reference simulation. On the other hand assuming a lower limit of $10\, {\rm M}_{\sun}$ (instead of $8\, {\rm M}_{\sun}$ ) results in an expected number of SNe of 3.9$\pm$2.1, which is $\sim$30\% lower than the expected rate obtained in the reference case.
 
\subsubsection*{Extinction}

For the reference case we assumed that the extinction scales with the SFR with a maximum value corresponding to the SFR peak $A_V=30$ mag.  
To test the uncertainty related to this assumption  we made two different tests. In one experiment we maintained the relation of $A_V$ with SFR but taking, alternatively, a peak extinction value  $A_V=10$ and $A_V=100$ mag.  The experiment gave  as expected number of SNe  $5.5\pm2.7$ and $4.7\pm2.3$, respectively. In the second experiment we assume that the extinction is constant through the galaxy and is $A_V=3.0$ mag. In this case the expected number is $5.3\pm 2.3$ identical to the value of the reference simulation.

The conclusion is that the uncertainty on the extinction does not affect significantly the simulation or, conversely, that  our experiment we cannot  probe  the extinction distribution.

\subsubsection*{Star Formation Distribution}

The spatial distribution of SFR is an important, and the most uncertain, ingredient of the simulation. For instance, if we assume that the SFR is confined in the very inner regions, say in the inner $3-500$ pc, the resulting SNe will remain unaccessible to our search. On the other hand, the fact that in some SBs the SFR is extended has been confirmed by different studies \citep[eg.][]{mcquinn:2012zr}, not to mention that many of the SNe we have discovered are at significant radial distances (cf. Tab.~\ref{snelist}).

As we described in Sect.~\ref{sfrdistribution} as proxy of the SFR distribution we use $L_K^\alpha$ where $\alpha$ range from 1 to 2 depending on the galaxy compactness class (Tab.~\ref{classification}).  To test for the uncertainties of this assumption we performed two simulations assuming that for all galaxy $\alpha$ is either 1 or 2. We obtained in the first case an expected rate of  $8.8\pm3.0$ and in the second case a value of $3.0\pm1.7 $. The latter occurs  because when the SFR is more concentrated, a large number of SNe remain hidden to our search due to the low search detection efficiency in the nuclear regions. 

The conclusion is that the uncertainty in the adopted SF distribution propagates with an error of 
$\sim 50\%$ on the expected SN number. We may consider that the actual good match of observations with  the reference simulation  argues in favor of the adopted prescription.

\section{Summary and Conclusions}\label{summary}

We have presented the analysis of an infrared SN search in a sample of 30 nearby SB galaxies, conducted   between 2009 and 2011, with the goal to verify the link between star formation and SN rate.  During our search we collected in total about 240 observations discovering  6 SNe, 4 of them with spectroscopic confirmation. 

How does this number compares with the expectation ?

Answering this question requires a detailed characterization of the SN search detection efficiency,  the galaxy properties (in particular SF rate and spatial distribution) and  the SN properties and progenitor scenarios. We included all these ingredients in a MonteCarlo simulation tool that, allowing for the stochastic nature of SN events, can be used to explore the distribution of the expected SN number and properties. 

First of all, we may remark that by itself the number of detected SNe is a proof of the high SFR in SBs. In fact if  we compute the expected number of SNe  in our survey based on the average SN rate per unit B luminosity or mass \citep{li:2011qf}, we would predict the discovery of 0.5 events (or more precisely,   50\% of the simulation predict the discovery of one event and none is expected in the other 50\%). The observed number is one order of magnitude higher, which is consistent with the fact that the TIR emission of SBs is about ten times higher than for normal SF galaxies with the same B luminosity. Indeed, it is well-known that the TIR luminosity is an excellent tracer for SFR, in particular in SBs. 

When we adopt the SFR from $L_{TIR}$  as input for the MonteCarlo experiment,  we find that the expected number of SNe in our search is $5.3\pm2.3$, SNe in excellent agreement with observations. In most cases we predict that only SN CC should be discovered while in the actual search we did detect one type Ia SN. Given that there is a sizable fraction of experiments (10\%) when this is predicted to occur we do not elaborate further  this issue.  Also, allowing for the low statistics, we find an excellent agreement between the predicted and observed SN properties, namely apparent magnitude at discovery, extinction and location inside the host galaxies.

We performed a number of tests to verify the dependence of the simulation outcomes from the input parameters. For the SN search characterization we show that an accurate estimate of the magnitude limit for SN detection is crucial. This is why we spend a considerable effort in artificial star experiments (possibly the single most expensive task of our project). For the galaxy characterization the most uncertain input is the SF spatial distribution. With some creativity, we devised a prescription that seems to work, but it is certain that this is a place for improvements when new, high resolution SB maps will become available.  Instead, we found that our results are not sensitive to the uncertainty on the amount of extinction because where extinction is very high (the dense SB regions) our search is limited by the bright magnitude detection limit. SNe in these regions remain hidden to our search almost independently on the amount of extinction. Based on our simulation we estimated that the fraction of hidden SNe is very significant, that is $\sim 60\%$  with an upper limit of 75\% if we account for the poissonian uncertainties in the number of detected events.
Finally, for the SN progenitor scenarios the larger uncertainty is the lower limit of the progenitor mass range. If we adopt a lower limit $M_{CC}^L = 10\, {\rm M}_{\sun}$ instead of  $8 \,{\rm M}_{\sun} $ as in the reference simulation, the expected number of SNe would be 30\% lower than observed.

Our results appear in good agreement with those of previous similar searches \citep[][cf. Sect.~\ref{intro}]{mannucci:2003wl,cresci:2007ly,mattila:2007bh,mattila:2012er}.  In broad terms, the overall conclusion of all these studies can be expressed as follows:  the number of (CC) SNe found in SBs galaxies is consistent with that predicted from the high SFR (and the canonical mass range for the progenitors) when we recognize that a major fraction of the events remains hidden in the unaccessible SB regions. As stressed by \cite{mattila:2012er}, this has important consequences for the use of SN~CC as probe of the cosmic SFR, because the fraction of SBs is expected to increase with redshifts \citep[cf.][]{melinder:2012ly,dahlen:2012zr}

While continuing to search for SNe in SBs, in optical and infrared, can certainly help to improve the still low statistics, one may argue at this point for a change of strategy.

In this respect good example is the attempt to reveal some of the hidden SN~CC through infrared SN searches which exploits  adaptive optics at large telescopes, eg. Gemini or VLT.  The results are encouraging with the discovery of two SNe with very high extinction, namely SN~2004ip with $A_V$ between 5 and 40 mag \citep{mattila:2007nb} and  SN~2008cs with $A_V~16$ mag  \citep{kankare:2008ys}, though we may notice that both objects were too faint for spectroscopic confirmation.  Other two SNe were  discovered very close to the galaxy nucleus, namely SN~2010cu at a radial distance of 180 pc and SN~2011hi at 380 pc \citep{kankare:2012zr}, though in these cases the low extinction suggests that the low radial distance is a projection effect (also in these cases no spectroscopic classification was obtained).  The extinction towards SN 2011hi was revised by \cite{romero:2012rc} using Gemini-N data. They demonstrate that this is most likely a SN IIP with $A_V$ of 5-7 mag. Because of the need to monitor one galaxy at the time and to access heavily subscribed large telescopes, this approach will not result in large statistics though even a few events may be molstly valuable to explore the very obscured nuclear regions.

On the other hand, a  new opportunity  that should be explored is the piggy-back on wide field extragalactic surveys of the next generation infrared facilities, in particular EUCLID. 
This would allow for the first time to perform IR SN searches on large sample of galaxies exploring a range of SF activity and, by monitoring galaxies at different redshifts, probe the cosmic evolution.

\begin{acknowledgements}
We thank the referee, Seppo Mattila, for the careful reading and the very useful comments.\\
We particular thank  Anna Feltre (ESO), for her help inestimating   the possible contribution by AGNs to the FIR luminosity of the galaxies and to  Barbara Lo Faro (Astronomy Department of Padova) for helpful discussions and suggestions.\\
We acknowledge the support of the PRIN-INAF 2009 with the project "Supernovae
Variety and Nucleosynthesis Yields".\\
E.C., L.G., S.B., A.P. and M.T. are partially supported by the PRIN-INAF 2011 with
the project "Transient Universe: from ESO Large to PESSTO".\\
N.E.R. acknowledges financial support by the MICINN grant AYA08-1839/ESP, AYA2011-24704/ESP, and by the ESF EUROCORES Program EUROGENESIS (MINECO grants EUI2009--04170).\\
F.B. acknowledges support from FONDECYT through Postdoctoral grant 3120227 and
from the Millennium Center for Supernova Science through grant
P10-064-F (funded by ''Programa Bicentenario de Ciencia y Tecnologıa
de CONICYT'' and ''Programa Iniciativa Cientifıfica Milenio de
MIDEPLAN'').

\end{acknowledgements}



\bibliographystyle{aa}
\bibliography{hawkiSB_EC}

\end{document}